\documentclass[10pt,journal,final,twocolumn]{IEEEtran}
\usepackage{graphicx} % for pdf, bitmapped graphics files
\usepackage{epsfig} % for postscript graphics files
\usepackage{mathptmx} % assumes new font selection scheme installed
\usepackage{times} % assumes new font selection scheme installed
\usepackage{amsmath} % assumes amsmath package installed
\usepackage{amssymb}  % assumes amsmath package installed
\usepackage{color}
\usepackage{authblk}
\usepackage[caption=false,font=footnotesize]{subfig}
\usepackage{setspace}
\usepackage{algorithm}
\usepackage{algorithmic}
\DeclareMathOperator*{\argmax}{argmax}

\newtheorem{theorem}{Theorem}
\newtheorem{lemma}{Lemma}

\newtheorem{definition}{Definition}

\newcommand{\paren}[1]{\left(#1\right)}
\newcommand{\sqparen}[1]{\left[#1\right]}
\newcommand{\brparen}[1]{\left\{#1\right\}}
\newcommand{\field}[1]{\ensuremath{\mathbb{#1}}}
\newcommand{\N}{\ensuremath{\field{N}}} % natural numbers
\newcommand{\R}{\ensuremath{\field{R}}} % real numbers
\newcommand{\Rp}{\ensuremath{\R_+}} % positive real numbers
\newcommand{\PRP}[1]{\ensuremath{\mathsf{Pr}\left(#1\right)}} %\probability with parentheses
\newcommand{\vecbold}[1]{\ensuremath{\boldsymbol{#1}}}

\newcommand{\E}[1]{{\mathsf{E}}\left[{#1}\right]}

\newcommand{\Prob}[1]{\mathsf{Pr}\left\{ {#1}\right\}}

\newcommand{\Qv}{{\boldsymbol{Q}}}

\renewcommand\boldsymbol[1]{\pmb{#1}}

\begin{document}
%\onehalfspacing
\title{{Stability and Dynamic Control of Underlay Mobile Edge Networks}}%{\thanks{This material is based upon work supported by the National
%Science Foundation under Grants CNS-0831919, CCF-0916664, CAREER-1054738.}}\thanks{Portions of this work were presented at Asilomar Conference on Signals, Systems, and Computers (Asilomar '10), Pacific Grove, CA.}}
% author names and affiliations
% use a multiple column layout for up to three different
% affiliations
%\author{\IEEEauthorblockN{C.~Emre Koksal}
%\IEEEauthorblockA{Department of Electrical and Computer Engineering,\\
%The Ohio State University\\
%Columbus, OH \\
%Email: koksal@ece.osu.edu} \and \IEEEauthorblockN{Ozgur Ercetin and
%Yunus Sarikaya}
%\IEEEauthorblockA{Faculty of Engineering and Natural Sciences,\\
%Sabanci University,\\
%Istanbul, TR.\\
%Email: \{oercetin,ysarikaya\}@sabanciuniv.edu}}

\author{{ Yunus Sarikaya ~\IEEEmembership{Member,~IEEE}, Hazer Inaltekin ~\IEEEmembership{Member,~IEEE}, Tansu Alpcan ~\IEEEmembership{Senior Member,~IEEE} and Jamie Evans ~\IEEEmembership{Member, ~IEEE}}%
\thanks{Y. Sarikaya, T. Alpcan, J. Evans (\{yunus.sarikaya, tansu.alpcan, jse\}@unimelb.edu.au) are with the Department of Electrical and Electronic Engineering, University of Melbourne, Australia}  \thanks{H. Inaltekin (hinaltek@princeton.edu)  is with the Department of Electrical Engineering, Princeton, NJ 08544, USA.}}%
%\thanks{C.~E. Koksal (koksal@ece.osu.edu) is with the Department of Electrical and Computer Engineering at The Ohio State University, Columbus, OH.}  \thanks{O. Ercetin (email: oercetin@sabanciuniv.edu) and  Y.Sarikaya (email: sarikaya@su.sabanciuniv.edu) are with the Department of Electronics Engineering, Faculty of Engineering and Natural Sciences, Sabanci University, 34956 Istanbul, Turkey.}}

\maketitle

\begin{abstract}
 This paper studies the stability and dynamic control of underlay mobile edge networks. %subject to quality-of-service requirements at the network core. 
First, the stability region for a multiuser edge network is obtained under the assumption of full channel state information. This result provides a benchmark figure for comparing performance of the proposed algorithms. Second, a centralized joint flow control and scheduling algorithm is proposed to stabilize the queues of edge devices while respecting the average and instantaneous interference power constraints at the core access point. This algorithm is proven to converge to a utility point arbitrarily close to the maximum achievable utility within the stability region. Finally, more practical implementation issues such as distributed scheduling are examined by designing efficient scheduling algorithms taking advantage of communication diversity. The proposed distributed solutions utilize mini-slots for contention resolution and  achieve a certain fraction of the utility optimal point. The performance lower bounds for distributed algorithms are determined analytically. The detailed simulation study is performed to pinpoint the cost of distributed control for mobile edge networks with respect to centralized control.

\end{abstract}
\begin{IEEEkeywords}
Mobile Edge Networks, Stability, Flow Control, Scheduling, Cross-layer Design.
\end{IEEEkeywords}
\section{Introduction}

\subsection{Background and Motivation}

The predictions of Cisco visual networking index indicate that mobile data traffic has grown $4,000$-fold over the past ten years \cite{Cisco16}. One main driver for such an unprecented growth is the surge of computationally powerful devices close to the network edge like smartphones, tablets, connected vehicles, smart meters, femtocells, wireless-enabled industrial robots/drones \cite{Chiang16, Kelly16,Mearian13}, to name a few. Consequently, there is an increasing tendency to perform communication and signal processing tasks at the network edge in next generation LTE-A networks through the enablement of technologies such as fog computing, WiFi direct, D2D/IoT communication, cognitive radio and femto-cells \cite{Chiang16, Akyildiz16, Yanikomeroglu14}. This paradigm shift in networking automatically triggers the need for a thorough investigation of multi-tier network architectures in which multiple network tiers with different underlying technology components can co-exist together in the same spectrum. 

The most prominent advantage of mobile edge networking is to ease the communication and computation burden on the core network by making use of the immersive distributed network of devices at the network edge. As such, data generated at the edge stays at the edge for communication and computation purposes such as stream mining and embedded artificial intelligence, without a need to traverse the core network anymore. This approach not only improves the efficiency of spectrum usage, but also has a great potential for enhancing the network performance expressed in terms of capacity, coverage, energy efficiency and end-to-end delay \cite{Lei12,Fodor12}.

The main challenge now is to adapt new approaches for networking at large so that direct communication among edge devices can seamlessly coexist with inbound and outbound data traffic from the core network in the same frequency band. As is common in the cognitive radio literature \cite{Goldsmith09}, two outstanding approaches for the coexistence of different technologies at the network edge and core can be conceived to be inband/outband underlay and overlay communications.  The main focus of this paper will be on the inband underlay communications in which edge devices utilize the same spectrum with the core access point (AP) in a two-tier network architecture.  In this setup, the radio frequency spectrum is the shared communication resource whose access must be regulated for network stability and performance optimization. This will be done at the network edge by designing smart interference management strategies and appropriate cross-layer resource allocation algorithms. %In particular, we will discover the maximum communication rates for which the queue sizes of edge devices stay finite with bounded and tolerable distortion to data traffic at the core AP.}  

%In the literature, various paradigms are introduced based on the spectrum in which heterogeneous communication occurs. These paradigms are classified as overlay and underlay inband and outband communications. The paradigm of interest in this paper is underlay inband communications, in which both micro-cell with low-power devices and macro-cell with high-power devices use the same spectrum to transmit their data. The existing researches allow micro-cell consisting of cognitive and/or D2D links as an underlay to the macro-cell network to increase the spectral efficiency [1, 3]. However, operating in the same licensed band, devices will inevitably impact macrocell users by causing interference. To ensure minimal impact on the performance of existing macro-cell BSs or access points (APs), a two-tier network needs to be designed with smart interference management strategies and appropriate resource allocation schemes.  } 

Important use cases of mobile edge networking include high data rate wireless services, IoT applications and industrial control systems \cite{Chiang16}. Many data-intensive services at the network edge such as virtual reality, online gaming, video sharing, vehicle-to-vehicle communication and proximity-aware social networks require small end-to-end delay of incoming data packets. The same also holds correct for many IoT applications and industrial control systems with strict deadlines on sensor-plant communication and control actions. Therefore, in addition to queue stability, an overall utility for communicating over the network edge must be maximized, as a measure of service satisfaction in different applications. To this end, we combine a variety of basic networking mechanisms such as flow control and scheduling in the context of underlay mobile edge networking. In particular, by modeling the entire problem as that of a network utility maximization, we develop the utility-optimal cross-layer dynamic flow control and scheduling mechanisms achieving the optimum utility point within the stability region subject to various interference power constraints at the core network. The main analytical tool used for this purpose is the stochastic network optimization framework put forward in \cite{Georgiadis}. The motivation to follow this approach is to investigate network stability and optimality jointly as is also done in \cite{kelly, Kar} to address fairness issues by investigating the scheduling problem and network utility maximization (NUM) together.      

\vspace{-0.0in}
\subsection{Contributions}

The main contributions of this paper are two-fold. First, a thorough analysis for interference-aware mobile edge networking is provided. In doing so, we derive the \textit{stability region} for edge devices operating under interference constraints at the network core. The notion of stability region here refers to the set of rates achievable by any feasible flow control and scheduling policies not violating predefined hard interference limitations at the core AP. The interference regulated stability region is  compared with the one without any QoS guarantees at the core network, which leads to the quantification of rate loss due to interference-aware operation of the edge network. 

We then formulate a resource allocation problem for interference-aware edge networking as a NUM problem, in which the optimal scheduling of edge devices is implemented at the MAC layer, while  the flow control is realized at the transport layer. We propose a cross-layer dynamic control algorithm for solving the scheduling and flow control problem jointly. It is shown that the proposed cross-layer design achieves a utility point arbitrarily close to the maximum achievable utility. In particular, the flow control algorithm moves the rate vector to the Pareto boundary of the stability region, %based on edge user utility functions, 
while the scheduling algorithm ensures that the core network constraint qualifications are met.

As a second contribution, we examine the problem of practical implementation of the above cross-layer design in the absence of a centralized scheduler. Specifically, we design simple but efficient distributed channel access algorithms, called \textit{channel-aware} distributed schedulers, where the edge devices decide to transmit or not based on their local information (i.e., their channels and queue backlogs). The proposed algorithms are channel-aware in the sense that they are able to take channel variations into account for scheduling decisions. 
%The proposed algorithm is based on \textit{carrier-sensing}, in which micro-cell users content the channel based on their local information, and if it senses another contention, it defers its transmission.  

We obtain analytical performance bounds on the dynamic control of the edge network based on the proposed channel-aware distributed schedulers. Now, the utility optimal point achieved through a centralized scheduler can no longer be guaranteed due to availability of limited information in the distributed mode of operation. To quantify the performance loss, we show that we can achieve an $\alpha^*$-fraction of the utility optimal point and obtain an analytical characterization of the parameter $\alpha^* \in \sqparen{0, 1}$. $\alpha^*$ can be adjusted as a function of the contention level and the number of mini-slots used by the distributed scheduler to resolve contention. We demonstrate the advantages of the proposed distributed solutions by means of an extensive simulation study.
%\item[{\bf (c)} ] We show that the proposed optimal scheme has high
%complexity in terms of computational overhead.  Thus, we design a
%sub-optimal solution based on the idea that decoupling the part of
%the problem which imposes high complexity, and solving this part
%offline.
%\item[{\bf (c)} ]  We obtain a closed form solution of optimal jamming powers when the instantaneous channel state of the eavesdropper channel is not available.
%Such a closed form solution is non-existent in the literature. In the
%literature, works either employ search algorithms to find the
%optimal solution or make assumptions on the probability of secrecy
%outages \cite{M_Ghaderi}, \cite{J_Huang}, which leads to high
%complexity in terms of computational overheard or a sub-optimal
%solution.
%\end{enumerate}

%We note that the proposed control algorithm is centralized.  

%D2D users may act autonomously only when the cellular infrastructure is unavailable.  

%throughput under the constraint of primary user's queue stability.

\vspace{-0.1in}
\section{Related Work}

%ADD Coginite radio network liteture and cut D2D communication literature.

Our results in this study crosscut a wide range of literature including mobile edge networking, D2D/IoT communication, cognitive radio networks and multi-tier HetNets. Hence, we will only mention the papers that are most relevant to ours, mostly focusing on the literature in underlay D2D communication and cognitive radio networks.

The papers such as \cite{Yu09,Yu11,Janis09,Min11,Xiao11, Hakola10, Jung12} investigate the resource allocation problem for underlay D2D/IoT networks. In \cite{Yu09, Yu11}, the authors consider a single cell scenario with a cellular base-station (BS) and two D2D users sharing the same spectrum. Power control is exercised on the BS and the spectrum usage of the D2D pairs is optimized by considering sum rate as the objective function subject to energy/power constraints under non-orthogonal and orthogonal sharing mode. 
%The authors show analytically that an optimal solution can be given either in closed form or can be chosen from a set. 
In order to further improve the gain from
intra-cell spectrum reuse, properly pairing cellular and D2D users for sharing the same resources was studied in \cite{Janis09, Min11}. 

In particular, \cite{Janis09} considers the control of interference from D2D links to the cellular users by limiting the maximum transmit power of the D2D users. In \cite{Min11}, the authors employ the knowledge of spatial interference for D2D receivers to maximize network capacity with multiuser MIMO. Cellular users in the vicinity of the interference-limited area are not scheduled. The optimal power allocation problems for D2D networks are analyzed in \cite{Xiao11, Hakola10, Jung12}. The authors in \cite{Xiao11} and \cite{Hakola10} show that the problem of optimal power allocation and mode selection are not tractable. They propose an alternative {\em greedy} heuristic algorithm to lessen interference at the core cellular AP. The proposed scheme is practical but cannot prevent excessive signaling overhead. The authors in \cite{Jung12} propose a method to identify power efficiency for D2D communication, which is a function of transmission rate and power consumption of the devices.

The papers \cite{Feng13, Su13, Le12} focus on performance optimization of D2D networks subject to certain QoS constraints. The authors in \cite{Feng13} consider throughput-optimal resource allocation problem with minimum rate guarantees for both D2D and regular users. %A maximum weight bi partite matching based scheme is developed to select a suitable D2D pair. 
%In \cite{Zhang13}, they propose a graph-based resource allocation method. The resulting problem turns out to be NP-hard, and hence a suboptimal heuristic that accounts for interference and network capacity is obtained. %In their proposed graph, each vertex represents a link (D2D or cellular) and each edge connecting two vertices shows the potential interference the two links. 
The paper \cite{Su13} formulates the problem of maximizing the system throughput with minimum data rate requirements by means of the particle swarm optimization framework to obtain a solution. In \cite{Le12}, they formulate fair resource allocation for D2D networks as an integer programming problem, which is NP-hard. Hence, they propose a sub-optimal solution that captures the interplay between different elements of the optimization problem in different phases. D2D caching networks are investigated in \cite{Gitzenis13} and \cite{Ji16}. The authors in \cite{Gitzenis13} analyze the scalability of multi-hop wireless communications for the case of replicated content across the nodes.  The system model in \cite{Ji16} incorporates traditional microwave and millimeter-wave D2D links, and they show that in non-asymptotic regimes, the proposed D2D system model offers very significant throughput gains with respect to BS-only schemes.

The above papers indicate the potential of D2D networks to improve spectrum efficiency and data throughput at the network edge if the interference can be regulated at the core cellular networks. The main point of difference between the current work and those above is the cross-layer design approach developed in this paper to address queue stability, scheduling and flow control jointly. Different from them, we obtain a parametric characterization for the stability region of devices at the network edge under hard interference limitations at the network core. Further, we obtain a cross-layer scheduling and flow control algorithm maximizing the overall network utility within this stability region without violating interference limitations.    

%The solutions of these approaches and their derived sub-optimal heuristic can indeed improve the system performance with QoS constraints. However, they do not seem to be a good candidate for time-stringent application with limited computational capacity. Nonetheless, the authors  of \cite{Yu09} and \cite{Yu11} derived the closed-form solution, which reduces the complexity. But they only consider a scenario with a cellular user and D2D communication pair, which is not practical in reality. The algorithm derived in this paper, which is shown to be close to optimal solution, is a simple index policy, i.e., the centralized entity performs simple algebraic manipulations to obtain the solution in each time instant. Furthermore, we consider a general network model which contains arbitraty number of D2D communication pairs.

 %On the other hand, even though many works on cognitive radio networks focus on overlay communication and designing spectrum sensing algorithms, there are considerable amount of works on 
In addition to above work, our results in this paper are also related to those on resource allocation and opportunistic scheduling for underlay cognitive radio networks (CRNs) \cite{Inaltekin12,Inaltekin14, Inaltekin16, Tran_11,Kabiri_13,urgaonkar,Krikidis_10,Wang_14,Kompella_13,Ashour_15}. The authors in \cite{Inaltekin12} investigated the optimal power control and the resulting throughput scaling laws for underlay CRNs under average interference power constraints at the primary users. The same results are extended to the fully distributed case and partial cooperation case between secondary and primary networks in \cite{Inaltekin14} and \cite{Inaltekin16}, respectively. The papers such as \cite{Tran_11} and \cite{Kabiri_13} analyze the performance of underlay CRNs subject to instantaneous interference power constraints to optimize outage probability and some queuing performance
metrics. 

Opportunistic scheduling for CRNs is studied in \cite{urgaonkar}, where Lyapunov optimization tools are used to design flow control, scheduling and resource allocation algorithms. Explicit performance bounds are derived. %Using the technique of virtual queues, the joint problem of stabilizing the queues of secondary nodes in addition to satisfying long term constraint on the collision probability is transformed into a queue stability problem \cite{urgaonkar}. The channel model in \cite{urgaonkar} only considers path loss. In queue based systems, cooperation from SUs increases packet throughput of PU. As a result PU packet queue is emptied more often, providing more silent slots for SU to transmit. 
In \cite{Krikidis_10}, the authors analyzed stable throughput for a primary multi-access system, where a secondary user (SU) receives packets from two primary users (PU) and relays them using the  superposition coding technique when the primary slot is idle. The paper \cite{Wang_14} studies the tradeoff between packet delay and energy consumption in a cooperative cognitive network. One common cooperation method is cooperative relaying \cite{Kompella_13}. In cooperative relaying, the SU receives the failed PU packets and relays them to the primary destination on the next transmission opportunity. An additional relay queue is needed at the SU source for this purpose.
Ashour {\em et al.} in \cite{Ashour_15} proposed an admission control algorithm as well as randomized service at the relay queue and analyzed stable throughput in CRNs. One distinctive aspect of the current paper from the existing work in underlay CRNs above is a thorough analysis of the distributed implementation of scheduling and flow control algorithms in a mobile edge networking setting. An efficient distributed scheduler is designed explicitly, and the loss from such a distributed mode of operation is characterized analytically.

{\allowdisplaybreaks\vspace{-0.0in}
\section{System Model, Scheduling Policy and Edge Network Stability}
\label{sec:model}

In this section, we will introduce the details of our system model and the definitions of the main concepts that are used throughout the paper in relation to this model. %Note that, in the rest of the paper, we use macro-cell BS and access point (AP) terms interchangeably.

\begin{figure}
\centerline{ \includegraphics[width=3.5in]{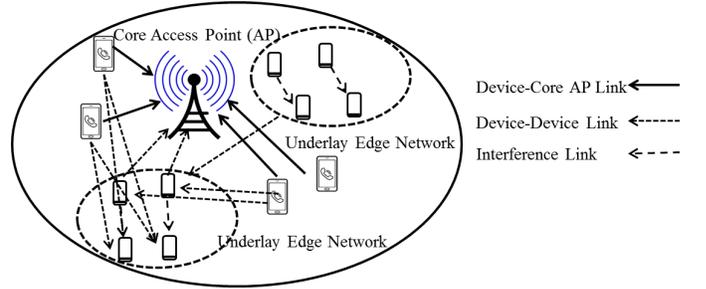}
} \caption{Network model consisting of $N$ edge device pairs that share a common frequency band with a core access point.}
\label{fig:d2dnetwork}
\end{figure}

\subsection{System Model} 

Our primary aim is to propose an efficient {\em cross-layer} design for mobile edge networks that lead to optimum utility point stabilizing the queues at edge devices. Both centralized and distributed approaches are investigated, with the centralized cross-layer algorithm (i.e., joint flow control and scheduling) being the benchmark to evaluate the performance of distributed algorithms. To this end, we consider a group of edge devices forming $N$ distinct links and sharing the same frequency band with a core network AP, as shown in Fig. \ref{fig:d2dnetwork}. Note that these communication links can be considered to be among D2D pairs, cognitive devices or femtocell users at the network edge depending on the application scenario. The edge devices are in close proximity of each other so that they can reach to their intended receivers in a single hop, yet they cause excessive interference to each other when multiple pairs are active at the same time. This leads to a fully connected interference graph topology with collision model for the edge network.

%characterize the maximum rates of communication that can be {\em stably} supported by a D2D %network and to discover the cross-layer {\em centralized} and/or {\em distributed} control %mechanisms (i.e., flow control and scheduling) that can achieve these rates with provable performance guarantees.  
%, i.e., bits over a link can be decoded successfully only when there is {\em one} active D2D %pair.          

%We consider an underlay radio network in which $N$ distinct D2D pairs share a frequency  band with the access point (AP) as shown in Fig. \ref{fig:d2dnetwork}. Each pair has data to be transmitted and its transmissions interfere with the signal
%reception at the AP. We assume fully connected interference graph with collision model, i.e., if multiple links are active at the same time instant, packet collisions take place and no data is transmitted over any link.

The devices operate in slotted time with slot indices represented by $t \in \N$. The link qualities vary over time according to the {\em block} fading model, in which the channel gain is constant over a time slot and changes from one slot to another independently according to a common fading distribution. We use $h_i(t)$, $i = 1, \ldots, N$, to represent the {\em direct} channel gain between the transmitter and receiver of the $i$th link. These direct channel gains are independent and identically distributed (iid) over users as well as over time. Operating in the same frequency band, the devices also cause interference to the core AP in Fig. \ref{fig:d2dnetwork}. We denote the {\em interference} channel gain between the transmitter of the $i$th edge link and the core AP by $g_i(t)$ for $i = 1, \ldots, N$. Furthermore, there may be other surrounding edge devices and core devices transmitting data to the core AP, not too close but causing some non-negligible interference to the edge network in question. We denote the total interference caused to the pair $i$ by $I_i(t)$ for $i = 1, \ldots, N$. Again, interference channel gains obey to the iid block fading model (possibly with a different distribution than that of the direct channel gains) as described above. We assume that the channel gains and inter-edge network interference levels are drawn from continuous distributions. For notational simplicity, we often use  the vector notation $\vecbold{h}(t) = \sqparen{h_1(t), \ldots, h_N(t)}$, $\vecbold{g}(t) = \sqparen{g_1(t), \ldots, g_N(t)}$ and $\vecbold{I}(t) = \sqparen{I_1(t), \ldots, I_N(t)}$ to denote the channel gains and interference values more compactly.  

For the sake of comprehending the interplay between the scheduling decisions at the MAC layer and the flow control decisions at the transport layer better, it is assumed that no power control is exercised at the physical layer of the network edge and all devices transmit at a constant power level $P$ over all time slots.  This assumption will help us to distill the effect of physical layer parameters on the interactions of the upper layer scheduling and flow control protocols, which is the main focus of the current paper.  In this setting, an important quantity of interest that determines the network performance is the rates (measured in units of bits/slot) offered over a communication link during time slot $t$. We assume that these communication rates are described by the functions $R_i(t)$ (as functions of transmission power levels, channel gains and interference levels) for $i=1, \ldots, N$. Even though we do not assume any specific functional form for $R_i(t)$, which is usually determined by the coding and communication technologies embedded in the transceiver circuits of the edge devices, we require that $R_i(t)$ has a bounded second moment, i.e., $\E{R_i(t)^2} \leq R_{max}^2$ for all $t \in \N$.\footnote{For example, if the Shannon capacity formula is used to quantify the communication rates for the $i$th link, $R_i(t)$ can be given as $R_i(t) = \log\paren{1 + \frac{P h_i(t)}{I_i(t) + N_0}}$, where $N_0$ represents the background noise power degrading transmissions over the $i$th link.} %Indeed, these communication rates are achievable by using Gaussian codebooks when slot durations are large enough \cite{Tse98}.} 
The significance of the rate function $R_i(t)$ in our analysis is that it will determine the {\em service} rates of the network layer queues maintained at the network edge.       

%Communication links vary over time according to independent and identically distributed (iid) block fading model, in which the channel gain is constant over a time slot and it is changing independently from slot to slot. We represent the direct channel gain in slot $t$ for the $i$th D2D link by $h_i(t)$, i.e.,  $h_i(t)$ is the channel gain between the transmitter of pair $i$ and the receiver of that pair. Similarly, the interference channel gain for the $i$th D2D communication link in slot $t$ is represented by $g_i(t)$, i.e., $g_i(t)$ is the channel gain between the transmitter of the $i$th link and the AP. We define the vector of channel gains by $\boldsymbol{h}(t) = [h_1(t), \ldots, h_N(t)]$ and $\boldsymbol{g}(t) = [g_1(t), \ldots, g_N(t)]$. We also denote the data rate of D2D pair $i$ in slot $t$ by $R_i(t)$. We assume that the transmit power of devices to be constant, identical to $P$ over all time slots $t$, i.e., $P_i(t) = P_j(t) = P, \ \forall j \neq i$. Based on coding/communication schemes, the channel rate is primarily determinant by the power level and channel state \footnote{ The maximum achievable rate defined as Shannon capacity is given as: $R_i(t) = \log(1 + Ph_i(t))$, where we assume that the cellular interference at the D2D links is normalized such that noise plus interference power at each D2D receiver is equal to unity. These communication rates are achievable by using Gaussian codebooks when slot durations are large enough. }.

An application runs at the application layer of each edge device, and generates the bits to be stored at the transport layer queues. These bits are accepted to the network layer according to a {\em flow control} mechanism that runs at the transport layer. We let $A_i(t)$ represent the amount of data (in bits per slot) that enters the network layer at the beginning of time slot $t$ and is stored at a network layer queue with size $Q_i(t)$ at edge device $i$, $i \in \brparen{1, \ldots, N}$. The relationship between these important network parameters at the queue level is displayed in Fig. \ref{fig:queue_model}.  

It is assumed that the input rate $A_i(t)$ is admissible in the sense that $A_i(t) \leq A_{max}$ for all $t \in \N$, and it has a long-term average $x_i$, i.e., $x_i = \limsup_{T \rightarrow \infty} \frac{1}{T} \sum_{t=1}^T A_i(t)$. The utility obtained by the communication over the $i$th  link, $U_i\paren{x_i}$, is a function of the long-term average rate $x_i$. We assume that $U_i(0) = 0$, and $U_i\paren{x}$ is a continuously differentiable, monotonically increasing and concave function of its argument. This concludes the description of our system model. %In the following subsections, we will formally introduce the notions of scheduling policy and network stability as well as providing some main definitions classifying/characterizing the scheduling policies and the edge network stability region.   

\begin{figure}
\centerline{ \includegraphics[width=3.0in]{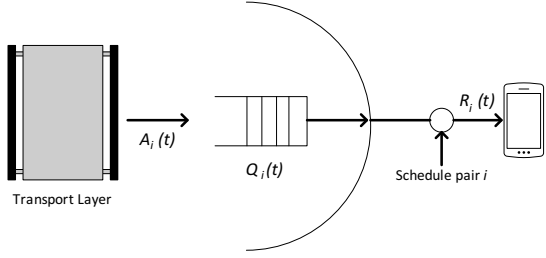}
} \caption{Queue model for edge pair $i$.}
\label{fig:queue_model}
\end{figure}

%We note that the service rate for a {\em stable} network layer queue  must be at least equal to $x_i$ \cite{Loynes62}, and that is why we %define the utility (or, satisfaction) obtained by the D2D pair $i$ from the transmission of its data as a function of $x_i$, since D2D pairs %cannot transmit more data than they already store in their queues.

%Let $A_i(t)$ represent the input rates in bits per channel use, at which data is injected in SU i in slot $t$. The rate $A_i(t)$ have long-term average $x_i$, i.e., $x_i = \lim_{T \rightarrow \infty} \frac{1}{T} \sum_{t=1}^T A_i(t)$. The utility obtained by D2D pair $i$ is a function of the long-term average rate $x_i$, since nodes cannot transmit more data than they receive. $U_i(x)$ represents the utility (or satisfaction) obtained by D2D pair $i$ from the transmission of its data, at a rate of $x$ bits per channel use.  We assume that $U_i(0) = 0$, and $U_i(\cdot)$ is continuously differentiable, monotonically increasing and concave function. The amount of traffic, $A_i(t)$ in the queues at D2D pair $i$ in slot $t$ is selected by D2D pair $i$ at the beginning of each block, and the injected information is stored in a queue with size $Q_i(t)$. The queue dynamics is illustrated in Fig. \ref{fig:queue_model}.

\vspace{-0.1in}
\subsection{Scheduling Policy}
Due to close geographical proximity of edge devices in our model, only one device pair in the edge network can communicate its data reliably over its respective wireless communication link. Hence, a scheduling decision must be made at the beginning of each time slot to select an {\em appropriate} user based on the current (both direct and interference) channel conditions.  For this purpose, roughly speaking, a scheduling policy should determine which set of links to be activated in each time slot for data transmission.  %Before giving the formal definition, let use define %$\Psi \in [0,1]^N$ is the set of all scheduling policies. 

\begin{definition}  \label{Def: Scheduling Policy} A scheduling policy $\vecbold{\cal I}: \Rp^{3N} \mapsto [0,1]^N$ is a vector-valued function $\vecbold{{\cal I}}\paren{\vecbold{h}(t),\vecbold{g}(t),\vecbold{I}(t)} = \sqparen{{\cal I}_1\paren{\vecbold{h}(t), \vecbold{g}(t),\vecbold{I}(t)}, \ldots, {\cal I}_N(\vecbold{h} (t),\vecbold{g}(t),\vecbold{I}(t)) }^\intercal$ that maps the direct and interference channel states to scheduling probabilities, i.e., ${\cal I}_i\paren{\vecbold{h}(t), \vecbold{g}(t),\vecbold{I}(t)} \in [0,1]^N$ for each $i \in \brparen{1, \ldots, N}$, and satisfies the feasibility constraint $\sum_{i=1}^N
{\cal I}_i\paren{\vecbold{h} (t),\vecbold{g}(t),\vecbold{I}(t)} \leq 1$.   
\end{definition} 

For ease of notation, we refer to ${\cal I}_i\paren{\vecbold{h} (t),\vecbold{g}(t),\vecbold{I}(t)}$ as ${\cal I}_i(t)$ in the rest of the paper. It should be noted that scheduling policies given in Definition \ref{Def: Scheduling Policy} constitute a collection of {\em randomized} control mechanisms for the edge network in question specifying scheduling probabilities for each pair of device.  Implicit in this definition, a scheduling policy does not allow two links to be active simultaneously due to the topological constraints of our network model. More explicitly, once scheduling probabilities are identified for all links in the edge network for time slot $t \in \N$, {\em only one} of them is selected for transmission by using the probability distribution induced by $\vecbold{{\cal I}}(t)$ over the set of edge device indices to determine the index of the selected link.  

An important subset of the randomized scheduling policies is the deterministic ones. We say that a scheduling policy $\vecbold{{\cal I}}(t) = \sqparen{{\cal I}_1(t), \ldots, {\cal I}_N(t) }^\intercal$ is a {\em deterministic} scheduling policy if ${\cal I}_i(t)$ is either zero or one for all $i \in \brparen{1, \ldots, N}$ and for all time slots $t \in \N$.  It will be shown below that the use of randomized scheduling policies will facilitate the mathematical analysis of the collection of optimization problems leading to the network stability region by turning them into convex optimization problems, {\em whilst} the solutions of these optimization problems lie in the set of deterministic scheduling policies.  

\vspace{-0.1in}
\subsection{Edge Network Stability}

In this part, we will provide the details of the notion of the edge network stability by relating the scheduling policies to the queue level dynamics of edge devices. To this end, we will first put forward the notion of interference-aware edge network operation.  All the network stability definitions presented afterwards will be with respect to this notion of interference-aware operation.    

The main communication paradigm of interest that we focus on for the coexistence of an edge network with core network users in the same spectrum is the {\em underlay} paradigm \cite{Goldsmith09}.  The main idea underpinning the underlay communication paradigm is that the edge network can utilize the same spectrum with the core AP as long as it does not cause harmful degradation to the data communication at the core AP by keeping the interference levels (instantaneous and average) below pre-specified interference threshold values. This leads to the interference-aware edge network operation, formally defined as below.     

%In this paper, we are interested in network stability in interference-limited network. Hence, we next give the definitions of interference-limited network, network stability and stability region.

\begin{definition} 
An edge network is said to be an {\em interference-aware} network if the average and instantaneous interference power levels that it causes to the core network AP is bounded above by the pre-specified interference threshold values as
\begin{eqnarray}
\sum_{i=1}^N\E{P g_i(t) {\cal I}_i(t)} \leq \gamma \mbox{ and } {\cal I}_i(t) = 0 \mbox{ if } Pg_i(t) > \nu, 
\label{eq:interference_const}
\end{eqnarray}
%and 
%\vspace{-0.0cm}
%\begin{eqnarray}
%\sum_{i=1}^N Pg_i(t){\cal I}_i(t) \leq \nu, 
%\label{eq:interference_const_inst}
%\end{eqnarray}
where $\gamma$ and $\nu$ denote the upper limits on the aggregate average and individual instantaneous interference powers from all links in the edge network, respectively. 
\end{definition}

This definition makes the coupling between the scheduling policies and the restrictions due to the interference-aware operation explicit.  In particular, the optimum scheduling policy achieving the maximum communication rates that can be stably supported by an interference-aware edge network must strike a balance between choosing the best link at the network edge and respecting the radio etiquette rules arbitrating the spectrum access rights between edge devices and the core AP.  The above interference constraints are primarily designed to safeguard two types of data traffic at the core network against the harmful edge network interference.  The average interference constraint in \eqref{eq:interference_const} is for delay-insensitive data traffic (e.g., text messaging) for which the messages are encoded and decoded over many time slots.  On the other hand, the instantaneous interference constraint in \eqref{eq:interference_const} is for delay-sensitive data traffic (e.g., video streaming) for which the messages are encoded and decoded over a single time slot. An edge network may not know the type of data traffic at the core AP at any given particular time, and hence must respect both constraints simultaneously.

Next, we formally define the concept of the stability of an interference-aware edge network. As is standard \cite{Georgiadis}, stability here refers to being long-term averages of expected queue sizes finite, i.e., $\limsup_{T \rightarrow \infty} \frac{1}{T} \sum_{t = 0}^{T-1} \E{Q_i(t)} < \infty$. %for all $i \in \brparen{1, \ldots, N}$. 
Further, we say that the edge network is {\em stable} under $\vecbold{{\cal I}}(t)$ if the network layer queues of all edge devices are stable. An important concept that expands upon the definition of network stability and relates the flow control and scheduling mechanisms for an edge network is the network stability region, which is defined as below.  
%\begin{definition} \label{Def: Network Stability}
%Consider an interference-aware D2D network operating according to a feasible scheduling policy $\vecbold{{\cal I}}\paren{\vecbold{h}(t),\vecbold{g}(t)}$ that satisfies the interference constraints in \eqref{eq:interference_const} and \eqref{eq:interference_const_inst}.  Then, we say the network layer queue $Q_i(t)$ at device pair $i$ is stable under  $\vecbold{{\cal I}}\paren{\vecbold{h}(t),\vecbold{g}(t)}$ if
%\begin{eqnarray}
%\limsup_{T \rightarrow \infty} \frac{1}{T} \sum_{t = 0}^{T-1} \E{Q_i(t)} < \infty. 
%\end{eqnarray}
%Further, we say that the D2D network is stable under $\vecbold{{\cal I}}\paren{\vecbold{h}(t),\vecbold{g}(t)}$ if the network layer queues of all devices are stable. 
%\end{definition}

%Throughout this paper, we restrict our attention to the strong stability definition given above, and often use the term stability to refer to strong stability.

%It is known(e.g., [26]) that the capacity region is given by
% = { |   0 and 9?? 2 Co(M), < ??}, (2)
%where Co(M) is the convex hull of the set of feasible
%schedules in M. 

%We use the indicator variable ${\cal I}_i(t)$ to represent
%the scheduler decision:
%\begin{equation}
%{\cal I}_i(t)=\begin{cases} 1, & \text{information transmitted from D2D pair $i$} \\
%0, & \text{otherwise}
%\end{cases}.
%\end{equation}

\begin{definition} \label{Def: Network Stability Region}
The {\em network stability region} of an interference-aware edge network, denoted by $\Lambda$, consists of all arrival rate vectors $\vecbold{x} = \paren{x_1, x_2, \ldots , x_N}^\intercal$ such that there exists a scheduling policy $\boldsymbol{{\cal I}}(t)$ satisfying the conditions below for all $i \in \brparen{1, \ldots, N}$ and $t \in \N$:

\vspace{-0.15in}
\small
\begin{align}
&\E{{\cal I}_i(t) R_i(t)} \geq x_i, \label{eq:stab_const} \\
&\sum_{i=1}^N\E{P g_i(t){\cal I}_i(t)} \leq \gamma, \label{eq:int_const} \\
&{\cal I}_i(t) = 0 \mbox{ if } Pg_i(t) > \nu, \mbox{and }\sum_{i=1}^N {\cal I}_i(t) \leq 1.   \label{eq:sched_const}
\end{align}
\end{definition}		

\normalsize
The constraints in \eqref{eq:int_const} and \eqref{eq:sched_const} are due to the interference-aware operation of the edge network and the feasibility condition.  The constraint in \eqref{eq:stab_const} is the classical necessity constraint for the queue stability describing the fact that the incoming rate to the network layer queues should be equal to or smaller than the outgoing service rate, which depends on the choice of the scheduling policy in our particular edge networking scenario \cite{Neely05}.\footnote{%Note that $\Lambda$ is the minimal set that contains all achievable arrival rates, i.e., no vector $\vecbold{x}$ outside $\Lambda$ can be stabilized by any feasible and interference-aware scheduling policy. 
Although not needed in our analysis, it can be shown that the stability region $\Lambda$ is a convex set by using the standard time-averaging arguments.}

 In the next section, we will obtain the Pareto boundary of the network stability region, where no feasible and interference-aware scheduling policy can stabilize the edge network when the arrival rates are beyond this boundary. This will provide a complete characterization of $\Lambda$. This characterization will be carried out under the full channel-state information (CSI) assumption.  Although helpful in understanding the maximum rates that can be stably supported by an edge network, such a characterization of the network stability region does not provide us with any insights regarding how to design dynamic control mechanisms achieving the rates in $\Lambda$.  

To resolve this drawback, we design a dynamic but centralized flow control and scheduling algorithm that achieves all the rates within the network stability region in Section \ref{control}. The scheduling part of the proposed algorithm provides design insights into how to construct a feasible and interference-aware scheduling mechanism for an edge network. In addition to stabilizing an interference-aware edge network, the proposed algorithm also maximizes the collective utility of the edge devices. The flow control part of the proposed algorithm provides design insights into how to construct flow control mechanism to maximize collective network performance. The distributed solutions achieving these desirable properties up to some performance gaps are given in Section \ref{sec:dist_scheduling_selective}.     

%When we analyze the stability region, we assume that every user has full CSI of its direct and interference channels, i.e., $h_i(t)$ and $g_i(t)$ are available to D2D pair $i$ at every block $t$. Then, we design a dynamic algorithm to achieve the rates within the stability region. Indeed, the proposed dynamic algorithm performs joint flow control and scheduling which not only stabilizes the queues of D2D network, but also maximizes the collective utility of D2D network.  We give centralized and distributed solutions to the problem, and derive performance bounds of the algorithms.

 %Thus, the achievable individual and sum rates we
%derive constitute upper bounds on the achievable rates with partial
%CSI, subject to the perfect privacy constraint.
%On the other hand, when we formulate our problem as that of network
%utility maximization problem, we only assume knowledge of
%instantaneous channel gains without requiring the knowledge of prior
%distribution of channel gains. %Hence, private encoding is performed
%over a single block length unlike the case when achievable rates are
%calculated. Clearly, privacy rate attained with this scheme is lower
%than the achievable rate obtained with full CSI.
}
{\allowdisplaybreaks\vspace{-0.0in}
\section{Stability Region for Interference-Aware Edge Networks}
\label{sec:ach_int_limited}
%Consider two users scenario in which two SUs are transmitting
%information over their uplink channels resulting interference to the
%CBS. Here, we consider two different network types as interference
%limited and power-interference limited networks. In interference
%limited network, SUs' transmissions are limited by an average
%interference constraint, i.e., we required that expected value of
%interference to the AP due to transmissions of SUs is below a given
%threshold, $\gamma$. In power-interference limited networks, SUs'
%transmissions are limited by both individual power constraints of
%SUs and an average interference constraint that they cause to AP.
%In both network types, we intend to study the effect of interference
%constraint on the achievable rates.

In this section, we derive the boundary of the stability region of an
interference-aware edge network such that any arrival rate vector outside the closure of the boundary is unattainable. Then, we analyze the effect of interference-aware communication on the network stability region by comparing the boundaries obtained with and without interference constraints.

 %for the cases when the scheduling is
%performed by a centralized entity (centralized algorithm) %or when
%the nodes decide whether to transmit or not based on their local
%knowledge (distributed algorithm). 
%In an interference limited network, D2D transmissions are limited by an average interference
%constraint at the AP, i.e., we require that the long-term average of interference to
%the AP due to transmissions of D2D pairs is kept below a given threshold,
%$\gamma$. 

We begin our analysis by computing the boundary of network stability region. This is equivalent to maximizing the average outgoing (service) rate achieved by edge device $i$ for given average rate values of other devices. Recall that the average arrival rate should be smaller or equal to the average service rate in a stable network. Hence, we say that an arrival rate $x_i$ of an edge device $i \in \brparen{1, \ldots, N}$ that is larger than this maximized service rate cannot be achieved. Before giving the mathematical description of the problem, the following remark is important. Since we assume that the channels are ergodic and stationary, we utilize the statistical averages in constructing the optimization problem. Hence, we ignore the time index for the sake of notational simplicity in this section. But in the next section, when we perform dynamic control, we again introduce time index back. Further, for notational convenience, let $\Psi$ be the set of all scheduling policies. Therefore, the aim is to maximize $\E{{\cal I}_i(\boldsymbol{h},\boldsymbol{g},\boldsymbol{I})R_i}$, associated with the point $\E{{\cal I}_j(\boldsymbol{h},\boldsymbol{g},\boldsymbol{I})R_j}
= \alpha_j, \forall j \neq i$, by solving the following optimization problem:

%Recall that ${\cal I}_i(\boldsymbol{h}(t),\boldsymbol{g}(t))$ is the scheduling decision for pair $i$ in slot $t$. 
%\subsection{Achievable Rate of Interference Limited Network}
 %The reason of using constant transmit
%power is to be able to make a fair comparison with the case without
%interference constraint, since the achievable rates with varying
%power are unbounded without power and interference constraint.
%Recall that ${\cal I}_i(t)$ as the indicator variable, which takes
%on a value 1, if D2D pair $i$ transmits its information over block $t$ and
%0 otherwise. Then, the average rate achieved by D2D pair $i$ is $R_i^\text{avg} =
%\E{{\cal I}_i(t)R_i(t)}$.  

% To find the pareto boundary of the stability region, we maximize the average rate achieved by a pair by fixing the average rates achieved by %other users. That is to say,
\vspace{-0.15in}
\begin{align}
    \max_{\boldsymbol{{\cal I}} \in \Psi} & \ \E{{\cal I}_i(\boldsymbol{h},\boldsymbol{g},\boldsymbol{I})R_i}\label{eq:obj_linear}\\
    \mbox{subject to}& \   \E{{\cal I}_j(\boldsymbol{h},\boldsymbol{g},\boldsymbol{I})R_j} \geq \alpha_j, \ \forall j \neq i \label{eq:obj_trans} \\
    & \E{\sum_{j=1}^N P{\cal I}_j(\boldsymbol{h},\boldsymbol{g},\boldsymbol{I}) g_j} \leq
    \gamma \label{eq:obj_int} \\
		&\hspace{-0.5in}{\cal I}_j(\boldsymbol{h},\boldsymbol{g},\boldsymbol{I}) = 0 \mbox{ if } Pg_j > \nu \mbox{ and } 		\sum_{j=1}^N
{\cal I}_j(\boldsymbol{h},\boldsymbol{g},\boldsymbol{I}) \leq 1,
\label{eq:sch_const}
\end{align}
where the expectations are over the joint distribution of the
instantaneous channel gains of direct and interference channels as well as the interference values. %Note that, here we consider that the scheduling decision can take a value %between 0 and 1, . Then, the scheduling decision corresponds to the probability of given the particular %scheduling decision. 
We solve the above optimization problem using the dual method that is particularly
appealing to our problem structure, whose solution is given in the next theorem. %Since our problem is a convex problem, it can be shown that %the optimal solution of the dual problem is exactly the same as the solution of the original problem, i.e., the duaility gap is zero %\cite{Boyd}.

%programming problem has a dual form. The original problem is sometimes called
%the primal problem. It can be shown that the optimal solution of the dual of any convex problem is
%exactly the same as the solution of the primal problem

\begin{theorem}
A solution of the optimization problem given in \eqref{eq:obj_linear}-\eqref{eq:sch_const} is equal to
\begin{align*}
{\cal I}_j^*(\boldsymbol{h},\boldsymbol{g},\boldsymbol{I})=\begin{cases} 1, & \text{ if  }W_j = \max_{k \in C} W_k \\
0, & \text{otherwise}
\end{cases}
\end{align*}
where $C = \brparen{j: W_j \geq 0, Pg_j \leq \nu}$, $W_j =  \lambda_j^*R_j - \mu^*P g_j$ for all $j \neq i$, $W_i =  R_i - \mu^*Pg_i$, and
$\lambda_j^*$ and $\mu^*$ are Lagrange multipliers associated with the rate and interference constraints in \eqref{eq:obj_trans} and \eqref{eq:obj_int}, respectively. % for which $\E{{\cal I}_j^*(\boldsymbol{h},\boldsymbol{g},\boldsymbol{I}) R_j} = \alpha_j$, for all $j \neq i$ and the constrain.

%is the value of $\lambda_j$ for which $\E{{\cal
%I}_j^*(\boldsymbol{h},\boldsymbol{g}) R_j} = \alpha_j$. %, since $\lambda^*\left(\E{{\cal I}_j(\boldsymbol{h},\boldsymbol{g})R_j} -
%\alpha_j\right)\geq 0$. 
%Similarly the value of $\mu^*$ is the value
%of $\mu$ for which $\E{P \sum_{l=1}^N {\cal I}_j(\boldsymbol{h},\boldsymbol{g})g_j} = \gamma$.

\label{thm:optimalscheduling}
\end{theorem}

\begin{IEEEproof} Please see Appendix A. \end{IEEEproof} 

%To be more formal, we can use functional optimization. Readers are referred to texts \cite{Luo05} for comprehensive results on functional optimization.

Theorem \ref{thm:optimalscheduling} gives us the optimal scheduling policy $\vecbold{{\cal I}}^*$ achieving $\E{{\cal I}_j^*(\boldsymbol{h},\boldsymbol{g},\boldsymbol{I}) R_j} = \alpha_j$ for all $j\neq i$. Then, the boundary of the stability region can be attained by varying $\alpha_j, \forall j\neq i$, and obtaining the points where the average rates of edge device $i$ are maximized. Another important point is that even though we state the optimization problem for the randomized scheduling policies, i.e., ${\cal I}_j(\boldsymbol{h},\boldsymbol{g},\boldsymbol{I})  \in [0,1 ]$, the optimal solution turns out to be a deterministic scheduling policy, i.e., ${\cal I}_j(\boldsymbol{h},\boldsymbol{g},\boldsymbol{I}) $ is either zero or one. In addition, observe that if the condition $P g_j > \nu$ or $W_j < 0$ for all edge devices, then the channel remains idle. % i.e., $\sum_{l=1}^N {\cal I}_l(\boldsymbol{h},\boldsymbol{g},\boldsymbol{I})  = 0$. 
The reason is that the channel conditions are not good enough to access the channel at the expense of the interference caused to the core AP.

%Observe that ${\cal
%I}_l(\boldsymbol{h},\boldsymbol{g}) = 1, {\cal I}_j(\boldsymbol{h},\boldsymbol{g}) =
%0, \ \forall j \neq l$ if the objective function on \eqref{eq:obj_fun_sep} is maximized for the
%transmission of $l$th D2D pair, or it will choose ${\cal
%I}_l(\boldsymbol{h},\boldsymbol{g}) = 0, \ \forall l$ otherwise. 

%The solution of above problem is on the boundary corresponding to binary scheduling policy. That is to say, $The last
%condition suggests that 

As indicated above, there may be time instants during which the channel remains idle in an interference-aware edge network to safeguard the core AP. This will result in a decrease in optimal rates due to under-utilization of the channel. Consequently, it leads to a contraction of the network stability region. To understand this phenomenon better, we also derive the optimum scheduling policy without interference
constraints, and compare the achievable rate regions in both cases with and
without interference constraints. Following similar arguments above, we have the following optimum scheduling problem

\vspace{-0.15in}
\begin{align}
    \max_{\boldsymbol{{\cal I}} \in \Psi} &\E{{\cal I}_i(\boldsymbol{h},\boldsymbol{g},\boldsymbol{I})R_i} \label{eq:obj_wo_int}\\
    \mbox{subject to } & \E{{\cal I}_j(\boldsymbol{h},\boldsymbol{g},\boldsymbol{I})R_j} \geq \alpha_j, \ \forall j \neq i \label{eq:obj_trans1} \\
& \sum_{j=1}^N
{\cal I}_j\paren{\boldsymbol{h},\boldsymbol{g},\boldsymbol{I}} \leq 1
\label{eq:obj_trans2}		
\end{align}
without interference constraints, whose solution is given by the next theorem.

\begin{theorem}
A solution of the optimum scheduling problem in \eqref{eq:obj_wo_int}-\eqref{eq:obj_trans2} is equal to
\begin{align*}
{\cal I}_j^*(\boldsymbol{h},\boldsymbol{g},\boldsymbol{I})=\begin{cases} 1, & \text{ if  } \lambda_j^*R_j > \lambda_k^*R_k,   \ \forall k \neq j  \\
0, & \text{otherwise}
\end{cases}
\end{align*}
\end{theorem}
where $\lambda_j^*$ is the Lagrange multiplier associated with the rate constraint in \eqref{eq:obj_trans1}.

\begin{IEEEproof} The proof follows the similar lines with the proof of Theorem \ref{thm:optimalscheduling}, and hence is skipped to avoid repetitions. \end{IEEEproof} 
\begin{figure*}
\centerline{ \subfloat[Network stability region for varying interference
parameter $\gamma$]{\includegraphics[width=3.0in]{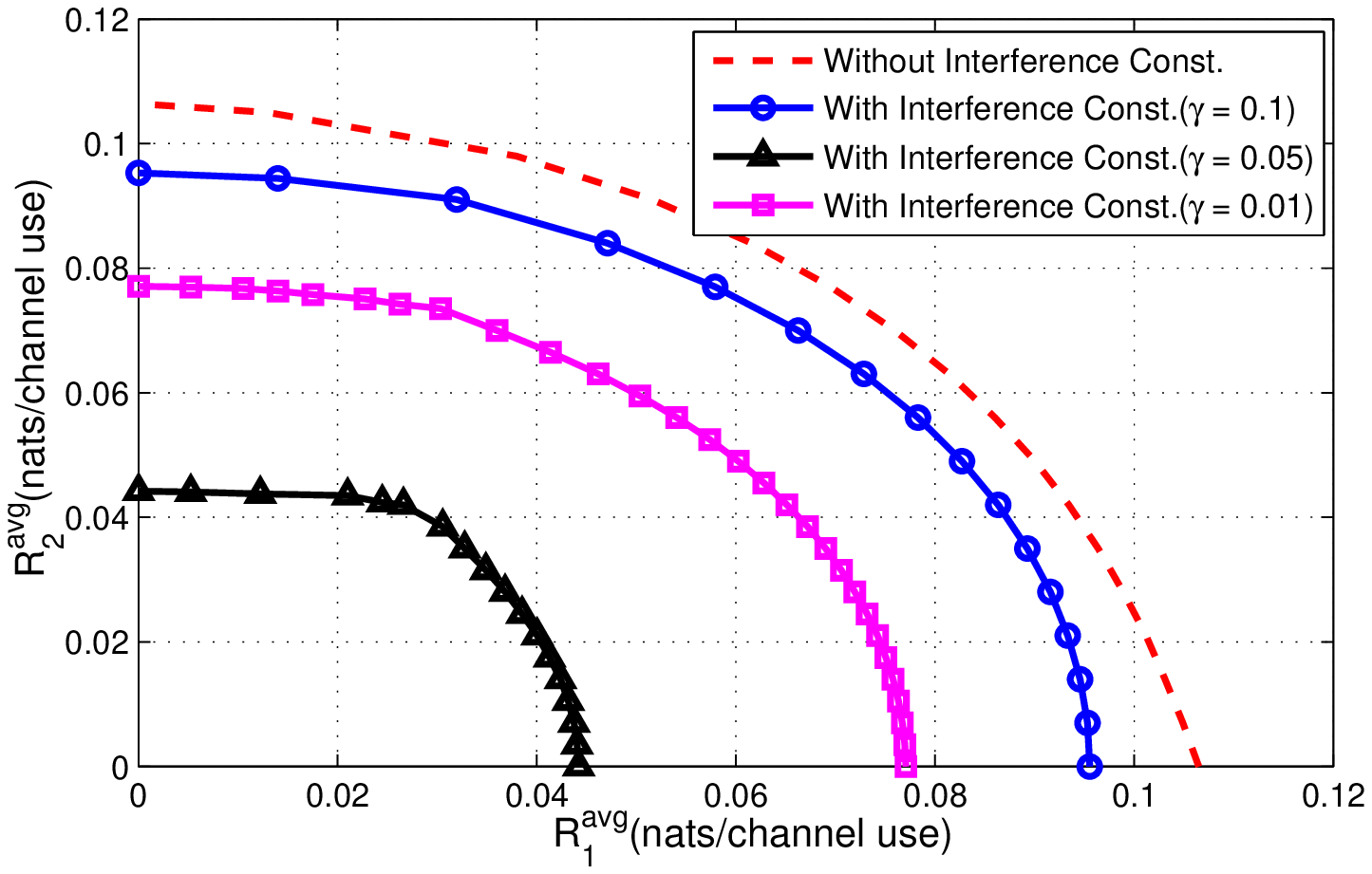}
\label{fig:gamma}} \subfloat[Network stability region for varying
interference channel gains.]{\includegraphics[width=3.0in]{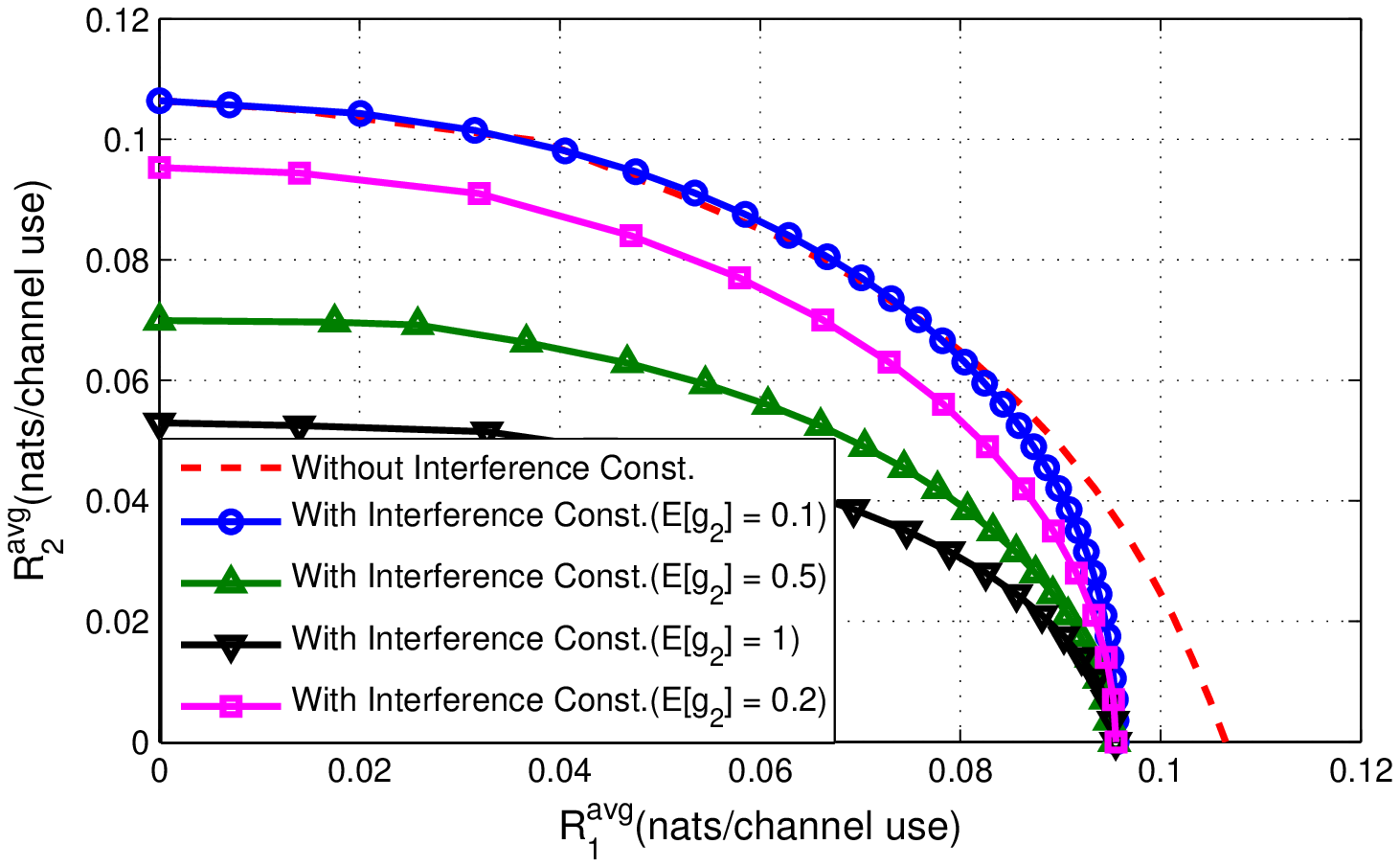}%
\label{fig:int_ch}  }} \caption{Network stability region for the communication scenario with two pairs of edge devices, where $R_i^{ \rm avg} = \E{{\cal I}_i^*\paren{\boldsymbol{h},\boldsymbol{g},\boldsymbol{I})}R_i}$. }
\end{figure*}
%Now, we consider a communication scenario containing only two edge pairs. In this case, the interference-aware optimum scheduling policy is given by 

%\vspace{-0.1in}
%\small
%\begin{align*}
%{\cal I}_1(\boldsymbol{h},\boldsymbol{g},\boldsymbol{I})=\begin{cases} 1, & \text{ if  } R_1 - \mu^*Pg_1 \geq \lambda^* R_2 - \mu^*Pg_2  \text{ and } R_1 > \mu^*Pg_1 \\
%0, & \text{otherwise}
%\end{cases}
%\end{align*}
%\normalsize
%and
%\small
%\begin{align*}
%{\cal I}_2(\boldsymbol{h},\boldsymbol{g},\boldsymbol{I})=\begin{cases} 1, & \text{ if  } R_1 - \mu^*Pg_1 < \lambda R_2 - \mu^*Pg_2  \text{ and } \lambda^* R_2 > \mu^*Pg_1 \\
%0, & \text{otherwise}
%\end{cases}.
%\end{align*}

%\normalsize

In Figs.~\ref{fig:gamma}-\ref{fig:int_ch}, the stability region for a two-link edge network is illustrated for Rayleigh fading direct and interference channels, in which the communication rate is selected as Shannon capacity, i.e., $R_i(t) = \log\left(1+\frac{Ph_i(t)}{N_0 + I_i}\right)$. Only average interference power constraint is considered for having clean exposition. We select 10 interfering links to the edge devices, and the mean channel gains of these links are randomly chosen  between $0.05$ and $0.2$ to simulate $\vecbold{I}$.\footnote{The total interference at the $i$th edge device pair, $I_i$, is the weighted sum of the gains of these interfering links, with weights being the transmission powers that are set to unity in Figs. ~\ref{fig:gamma}-\ref{fig:int_ch}.} To plot the stability regions, we varied the rate achieved by the second pair, calculated $\lambda^*$ and obtained 
the boundary rate pair for each point. Recall that $h_i$ and $g_i$ are the direct and interference channel gains of the $i$th pair, respectively. In Fig. \ref{fig:gamma}, we take
$\E{h_1}=\E{h_2} = 0.4$ and $\E{g_1}=\E{g_2} = 0.2$ to obtain
different boundary rate pair for varying interference parameter $\gamma$. %In simulations, we observe that the
%rate regions with and without interference constraint coincide for $\gamma = 0.25$. The
%reason is that the interference constraint in
%\eqref{eq:obj_int} becomes inactive in this case, i.e., $\mu^* = 0$. Thus, the
%optimal solutions for both problems are the same. 
As seen in Fig. \ref{fig:gamma}, when we decrease $\gamma$, i.e., the interference constraint is more stringent, the network stability region shrinks since both pairs have less transmission opportunities to meet the interference constraint. In Fig. \ref{fig:int_ch}, we fixed the value of $\gamma$ at $0.1$ and vary $\E{g_2}$. As seen in Fig. \ref{fig:int_ch}, when $\E{g_2} = 0.1$,  the network stability regions (with and without interference constraints) coincide for small rate values of edge user $1$, where the second pair takes a higher portion of transmissions. This observation results from the fact the interference constraint is inactive in this case and the second pair with smaller interference channel gain transmits predominantly. On the other hand, as $\E{g_2}$ increases, i.e., $\E{g_2}=0.2, 0.5$ and $1$, the interference experienced by the core AP starts to increase and the network stability region shrinks.

}
{\allowdisplaybreaks\section{Control of Underlay Edge Networks with Centralized Scheduling}
\label{control}

In the previous section, we characterize the stability region by obtaining maximum rates that an interference-aware edge network can support. In this section, we will present a dynamic control algorithm  that will solve a NUM problem while stabilizing the network layer queues in an edge network. To do so, we follow a cross-layer design approach. In the lower layer, the scheduling policy ensures network stability and satisfies the interference requirements. In the upper layer, on the other hand, flow control policy strives to move the network layer arrival rates to the optimal point within the stability region. Since the derived cross-layer algorithm will be a dynamic online algorithm, we will use the time index $t \in \N$ in this section again to indicate its operation in time.
%that opportunistically schedules the D2D pairs with the objective of
%maximizing total expected utility of the network subject to
%interference constraint to the AP %and individual power constraint
%of D2d pairs 
%while maintaining the stability of queues of D2D pairs.
%In the previous sections, the achievable rate region of the network
%is calculated based on the assumption that complete distributions of
%uplink and interference channel gains are available. In this
%section, we assume that the nodes are capable of obtaining perfect
%instantaneous CSI, but do not have access to the prior distributions
%of the channel states. 

The dynamic cross-layer algorithm takes the queue lengths (both virtual and real queues) and instantaneous direct and interference channel gains as input, and determines the scheduled device $i \in \brparen{1, \ldots, N}$ at each time slot as an output. We start our analysis by first formulating the NUM problem and providing the queue dynamics to set the stage for the cross-layer design approach. 

\subsection{NUM Problem Formulation}

Our objective is to stabilize the edge network while maximizing the sum of device utilities.  That is, we aim to find a solution for the following NUM problem:

\vspace{-0.15in}
\begin{align}
    \max_{\vecbold{x}} \ &\sum_{i=1}^N U_i(x_i) \label{eq:opt-objective} \\
    \mbox{ subject to } & \vecbold{x} \in \Lambda.  \label{eq:const-stability} %\\
    %& \E{P_i(t)} \leq \beta_i \ \forall i \label{eq:const-power}\\
    %& \sum_{i=1}^N \E{Pg_i(t)} \leq \gamma \label{eq:const-interference}
\end{align}

The objective function in \eqref{eq:opt-objective} accounts for the total expected utility of edge devices over random stationary channel conditions, interference values and scheduling decisions. The constraint in \eqref{eq:const-stability} ensures that network layer arrival rates of edge devices are within the rate region that can be stably supported by the edge network. The above problem could be in principle solved by means of standard convex optimization techniques if the stability region is known in advance. Although this approach may give us an idea about how to select transmission rates, it will not say anything about how we can reach the optimum operating point by relating the solution to the design of edge networks. Thus, in the following subsections, we develop a practical dynamic control algorithm to facilitate our understanding of the interplay between interference requirements and the critical functionalities of edge networks, such as scheduling and flow control.

\subsection{Queue Dynamics}

We assume that there is an infinite backlog of data at the transport layer of each edge device. Our proposed dynamic flow control algorithm determines the amount of traffic injected into the queues at the network layer. The dynamics of the network layer queue of the $i$th edge pair is given as

\vspace{-0.15in}
\begin{align}
Q_i(t+1)=\left[ Q_i(t)-{\cal I}_i(t) R_i(t)\right]^+ + A_i(t).
\label{eq:real_queue}
\end{align}

To meet the average interference constraint given in \eqref{eq:interference_const}, we also maintain a virtual queue

\vspace{-0.15in}
\begin{align}
Z(t+1) &= \left[ Z(t)- \gamma  + \sum_{i=1}^N {\cal I}_i(t)
Pg_i(t)\right]^+. \label{eq:interference_queue}
\end{align}

The state of the virtual queue at any given time instant is an indicator on the amount by which we exceed the allowable interference constraint. Thus, the larger the state of these queues, the more conservative our dynamic algorithm has to get towards meeting constraints, i.e., the less transmissions will take place by edge device pairs. The strong stability of virtual queues guarantees that the interference constraints are satisfied in the long run (i.e., see Theorem 5.1 in \cite{Georgiadis}).

%that yields a resulting matrix of throughput r is arbitrarily close to the optimal solution of ()-()

\subsection{Dynamic Control}

The proposed cross-layer dynamic control algorithm is based on the stochastic network optimization framework \cite{Georgiadis}. This approach enables us to obtain a solution for long-term stochastic optimization problems without requiring an explicit characterization of the stability region. Furthermore, it facilitates the simultaneous treatment of stability and performance optimization  by the introduction of virtual queues to transform performance constraints into queue stability problems. 

To this end, consider the queue backlog vectors for communication pairs, which are denoted as $\Qv(t)=(Q_1(t),\ldots, Q_N(t))$ and $Z(t)$. Let $L(\Qv(t),Z(t))$ be a quadratic Lyapunov function of real and virtual queue backlogs defined as:

\vspace{-0.15in}
\begin{equation} L(\Qv(t),Z(t)) =
\frac{1}{2} \left((Z(t))^2 +
\sum_{i=1}^N(Q_i(t))^2 \right).
\label{eq:lyapunov-function}
\end{equation}

Also, consider the one-step expected Lyapunov drift, $\Delta(t)$, for the Lyapunov function as \begin{equation*} \Delta(t) = \E{ L(\Qv(t+1),Z(t+1)) - L(\Qv(t),Z(t))|\Qv(t),Z(t)}.
%\label{eq:lyapunov-drift}
\end{equation*}

%Note that $\Delta(t) = 0$ if and only if all network queues are empty at
%time $t$, and that $\Delta(t)$ is large whenever one or more components in $\Delta(t)$ is large. 
The aim of the stochastic network optimization framework is to minimize the drift to ensure network stability, which can be achieved by having negative Lyapunov drift whenever the sum of queue backlogs is sufficiently large. Intuitively, this property ensures network stability because whenever the queue backlog vector leaves  the stability region, the negative drift eventually drives it back to this region. Furthermore, the following  utility-mixed Lyapunov drift
\vspace{-0.0in}
\begin{equation} \Delta^U(t)=\Delta(t) - V\E{\sum_{i=1}^N
U_i(A_i(t)) \ \Big | \ \Qv(t),Z(t)} \label{eq:deltawithreward}
\end{equation}
enables us to maximize the edge network performance in conjunction with the network stability, where the conditional expectation is taken with respect to all common randomness and $V>0$ is a design parameter. 

Next, we present the control algorithm that minimizes \eqref{eq:deltawithreward} and provide its optimality in Theorem \ref{thm:optimalcontrol}.

\noindent {\bf Control Algorithm:} Making an analogy to back pressure algorithm, we propose the following cross-layer algorithm that executes the following steps at each time $t \in \N$:
\begin{enumerate}
\item[\bf (1)] {\bf Upper Layer - Flow control:} The flow controller at each edge device observes its current queue backlog $Q_i(t)$. It then injects $A_i(t)$ bits, where $A_i(t)$ is the solution of the following optimization problem:
\begin{align}
A_i(t)=\argmax_{ 0 \leq x \leq A_{max} }\{ V U_i(x)- Q_i(t)x \}.
\end{align}
The design parameter $V > 0$ will determine the final performance of the proposed algorithm. The above identity involves maximizing a concave function, which can be easily solved by using  convex optimization techniques \cite{Boyd}.

\item[\bf (2)] {\bf Lower Layer - Scheduling:} A scheduler observes the backlogs in all edge devices and all fading/interference states. Then, it determines the scheduling decision for time slot $t \in \N$, $\boldsymbol{{\cal I}}(t)$, as the following index policy:

\begin{eqnarray*}
\setlength{\nulldelimiterspace}{0pt}
{\cal I}_i(t) = \left\{\begin{IEEEeqnarraybox}[\relax][c]{l's}
1, & if $i =\argmax_{j \in C} W_j(t)$\\
0, & otherwise
\end{IEEEeqnarraybox}\right. \label{Eqn: Centralized Scheduler}
\end{eqnarray*} 
for all $i \in \brparen{1, \ldots, N}$, where $C = \brparen{j: W_j \geq 0, P g_j(t) \leq \nu}$ and $W_i(t)$ is the weight of edge pair $i$ that is given as: 
\begin{align}
	W_i(t)= Q_i(t)R_i(t)-PZ(t)g_i(t).
	\label{eq:weight}
\end{align}
%D2D pair $i$ among the pairs that satisfies the instantaneous interference constraint, i.e., $Pg_i(t) \leq \nu$, and satisfy where
%\[ (\boldsymbol{{\cal I}}(t))=\argmax_{\boldsymbol{{\cal I}} \in \Psi }\ \left\{ \sum_i {\cal I}_i W_i(t)\right\}, \]
%\vspace{-0.15in}
%\begin{align*}
%\boldsymbol{{\cal I}}^*(t) =
%& \max_{\boldsymbol{{\cal I}} \in \Psi }\ \left\{ \sum_{i=1}^N {\cal I}_i W_i(t)\right\} \\
%\mbox{subject to } 	&\sum_{i=1}^N {\cal I}_i \leq 1 \mbox{ and } Pg_i(t) \leq \nu, \ \forall i,	
%\end{align*}
%\vspace{-0.1in}

%The above is a standard linear optimization problem, whose solution is obtained on the boundary. 
Specifically, among the edge pairs that satisfy the instantaneous interference constraint, the one having the maximum weight is allowed to transmit at a given time slot. If the set $C$ is an empty set, then no edge pair is scheduled for transmission. If the set $\argmax_{j \in C} W_j(t)$ is not singleton,  then any one of edge pairs in this set can be scheduled for transmission. For continuous interference channel states, $\argmax_{j \in C} W_j(t)$ is always singleton if there exists at least one element in $C$.   
\end{enumerate}

%we again show that the optimal scheduling policy is deterministic meaning that only one device is allowed to transmit at given time slot
 
We note that the parameter $V>0$ in the flow control algorithm determines the extent to which the utility optimization problem is emphasized. Indeed, if $V$ is large relative to the current backlog in the source queues, then the admitted rates $A_i(t)$ will be large, increasing the time average utility while consequently increasing the congestion level at the network edge. This effect is mitigated by more conservative flow control decisions as the backlog grows at the source queues. Note that the flow control algorithm is decentralized because the control values for each device require only knowledge of the queue backlogs at edge device pair $i \in \brparen{1, \ldots, N}$. 

In the scheduling policy, the weight equation \eqref{eq:weight} consists of a reward term $Q_i(t) R_i(t)$ and a cost term $P Z(t) g_i(t)$. Specifically, the larger the data queue backlog size $Q_i(t)$ and/or higher the instantaneous channel rate $R_i(t)$, the more likely the transmission from edge pair $i$ occurs. On the other hand, the larger the interference queue backlog size $Z(t)$ (representing the interference level caused to the core AP)
and/or higher the interference channel gain $g_i(t)$, the less
likely the transmission of edge pair $i$ takes place. In this setting, the flow control algorithm strives to maximize collective network utility, whereas the scheduling policy makes sure that the utility maximizing operating point is within the stability region. Indeed, by utilizing the proposed scheduling algorithm, we can achieve any point in the stability region.

\begin{theorem}
\label{thm:optimalcontrol}
Suppose $\boldsymbol{x^*} = [x_1^*, \ldots, x_N^*]$ is the average arrival rates produced by the proposed dynamic control algorithm. Then, for any flow parameter  $V > 0$,  the dynamic control algorithm yields the following performance bound for the aggregate network utility:
 \begin{align*}
    \sum_{i=1}^N  U_i(x_i^*) &\geq U^* - \frac{B_1}{V}
		\end{align*}		
while bounding the total long-term expected queue lengths as:	
\begin{align*}
    \limsup_{T \rightarrow \infty} \frac{1}{T} \sum_{\tau =
    0}^{T-1}\sum_{i=1}^N \E{Q_i(\tau)} &\leq
    \frac{B_1+V\kappa}{\epsilon_1},
 \end{align*}
 where $B_1,\epsilon_1,\kappa > 0$ are constants and $U^*$ is the optimal
 aggregate utility of the problem in
 \eqref{eq:opt-objective}-\eqref{eq:const-stability}. 
\end{theorem}

\begin{IEEEproof}  We omit the proof due to space limitation. See \cite{tech_report_D2D} for details. \end{IEEEproof}

%According to Theorem \ref{thm:optimalcontrol}, there is a trade-off in choosing the
%parameter V, i. e., larger values achieve a solution closer to
%the optimal, but at the same time increases the aggregate queue
%length. 

This theorem shows that the proposed dynamic control gets arbitrarily close to the optimal utility by choosing $V$ sufficiently large at the expense of proportionally increased average queue sizes. We note that the proposed dynamic control algorithm is not distributed since its scheduling part depends on global queue length information. As compared to the distributed scheduling algorithms, the centralized scheduling schemes usually lead to a better performance at the cost of requiring a central authority to allocate the network resources. In edge networking, such a central authority does not always exist. Furthermore, implementation of the centralized algorithms results in high overhead on the network due to the process of collecting channel conditions and queue states of all edge devices. In the remainder of the paper, we focus on designing distributed scheduling algorithms relaxing the assumptions necessary for the centralized algorithm. Note that the flow control part of the proposed solution is already distributed, i.e., each node decides its admitted flow based on only local information. Thus, it remains the same below.
}
{\allowdisplaybreaks\section{Channel-Aware Distributed Algorithms for Edge Networks}
\label{sec:dist_scheduling_selective}

In this section, we relax the requirement of having a central authority for scheduling edge devices in Section \ref{control} by investigating contention based  distributed scheduling algorithms with multiple round contention. The proposed algorithm will be called Channel-Aware Distributed Scheduler (CADS) that operates based on the local queue size and channel state information at each edge device pair. The distributed mode of operation necessitates the modification of the NUM problem as

%For that reason, a distributed scheduling
%algorithm would be very appealing even if it could only obtain a
%certain fraction of the achievable rate region of a centralized
%scheduling scheme. %Hence, in this section, we aim to design practical distributed algorithms with good performance guarantees.

%Due to limited information about the state of the network, obtaining the maximum weight at each time slot may not be feasible due to limited information, resulting in imperfect (sub-optimal) solution. This leads to reduction in stability region, i.e., distributed scheduling can only stabilize the rates within some portion of stability region. That is why, we modify the optimization problem for distributed scheduling as:
\vspace{-0.15in}
\begin{align}
    \max \ &\sum_{i=1}^N U_i(x_i) \label{eq:opt-objective_dist} \\
    \mbox{ subject to } & \vecbold{x} \in \alpha \Lambda, \label{eq:const-stability_dist} %\\
    %& \E{P_i(t)} \leq \beta_i \ \forall i \label{eq:const-power}\\
    %& \sum_{i=1}^N \E{Pg_i(t)} \leq \gamma \label{eq:const-interference}
\end{align}
where $\alpha$ is a contraction coefficient. The constraint in \eqref{eq:const-stability_dist} suggests that the distributed scheduling algorithms can still stabilize the edge network, provided that the arrival rates are interior
to $\alpha\Lambda$, which is an $\alpha$-scaled version of the stability region. In the remainder of the section, we will provide the details for CADS. % and obtain its performance bounds by characterizing $\alpha$.  
For analytical purposes, we assume that all direct and interference channel states are iid. However, we perform simulations for iid and non-iid cases and observe that the proposed algorithm often achieves scheduling performance closer to the centralized case. %far better than the obtained performance {\em lower} bounds. 
Furthermore, we will only assume the average interference constraint, but it is straightforward to incorporate the instantaneous interference requirement in the solution as well.

%a distributed scheduling algorithm with single round contention, called Interference Regulated Distributed Scheduler (IRDS), and obtain its %performance bounds by characterizing $\alpha$. IRDS  has low complexity but lacks exploiting diversity gain of the fading channels. In the next section, we will introduce a distributed scheduler with multiple discrete contention periods that is capable of exploiting the diversity gain.

\vspace{-0.15in}
\subsection{Contention Resolution Phase in CADS}

\begin{figure*}[htp]
\centerline{ \subfloat[Successful Contention ]{\includegraphics[width=3.0in]{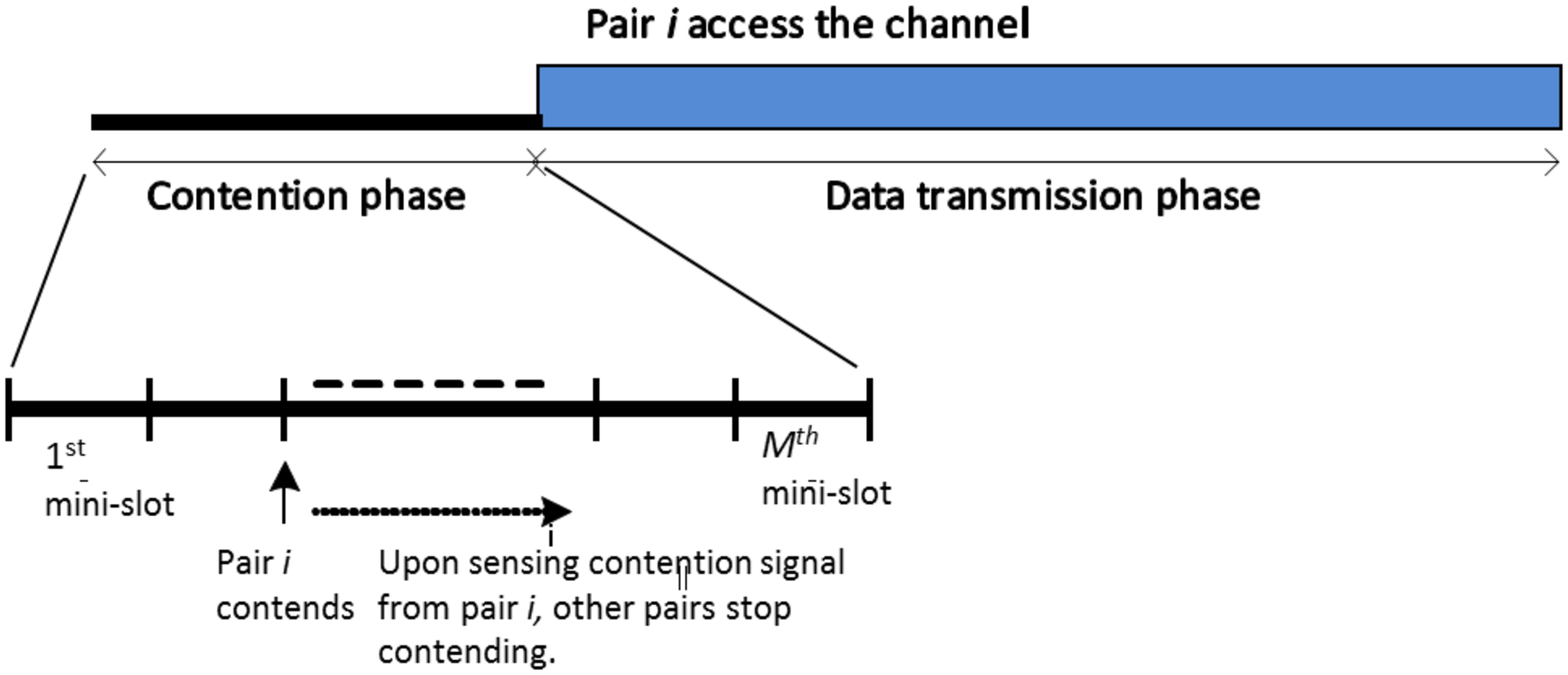}
\label{fig:cont1}} \\ \subfloat[Collision in Contention Phase]{\includegraphics[width=3.0in]{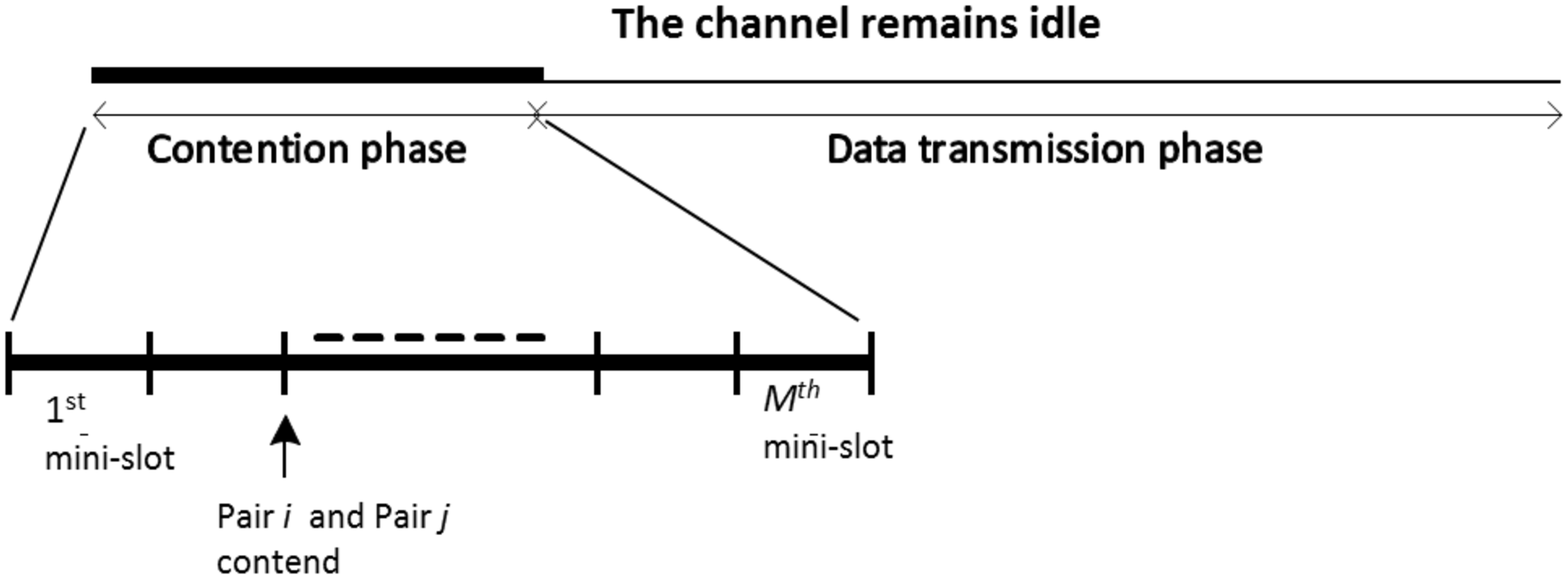}%
\label{fig:cont2}  }} \caption{An illustration for the steps in Algorithm \ref{alg:CSMA_Based}}
\end{figure*}

%In the previous section, we showed that SRC  algorithm can not take advantage of diversity gain of wireless fading channels. 
%In this part, we introduce the common operational principles shared by all CADS-type distributed schedulers that will be introduced in the subsequent sections. 
Operation of CADS takes place in slotted time in two phases: (i) contention phase and (ii) data transmission phase. The contention phase is composed of $M$ mini-slots, each of which is of enough duration to detect contention signals from other edge devices, i.e., a mini-slot must be at least $8 \mu\mbox{s}$ in an IEEE 802.11b environment. If $\tau$ is the ratio of the mini-slot duration to the duration of a regular time slot, then the parameter $M\tau$, which will appear below, signifies the fraction of time spent to resolve collisions. % by means of implementing a contention resolution phase before the transmission of actual data.   

The contention from edge devices for time slot $t \in \N$ is resolved as follows. The $i$th pair selects a mini-slot $m \in \brparen{1, \ldots, M}$ to send its contention signal.  The selected mini-slot $m$ depends on the pair $i$'s weight $W_i(t)$ that incorporates queue backlog, direct channel and interference channel information into a single parameter. If the pair $i$ senses a contention signal from another pair before the $m$th mini-slot, it stops contending for the channel and defers its data transmission to the next time slot, i.e., ${\cal I}_i(t) = 0$. Otherwise, it sends a contention signal in the beginning of the $m$th mini-slot. If no collision is sensed, then the $i$th pair obtains the access for the channel to transmit its data for the remaining part of the time slot, called the data transmission phase, commencing after the contention phase. If a collision is sensed, then the time slot remains idle and no data transmission takes place. These steps are visually illustrated in Fig. 4 and summarized in Algorithm \ref{alg:CSMA_Based} below.

%Secondly, we assume that when a pair detects a contention from other edge devices, it stops contenting the channel, i.e., the first pair which contends for the channel, obtains the right to access the channel and start transmitting in data transmission phase as illustrated in Fig. \ref{fig:cont1}. Furthermore, if more than one user contend in the same mini-slot, we say that collision takes place and the channel remains idle in data transmission phase as illustrated in Fig. \ref{fig:cont2} to prevent redundant interference to the core AP. The basic operational steps of CADSs is summarized in Algorithm \ref{alg:CSMA_Based}.

\begin{algorithm}
\caption{Contention Resolution in CADS}  
	1. At the beginning of each time slot $t \in \N$, the pair $i \in \brparen{1, \ldots, N}$ picks a mini-slot $m \in \{ 1, \ldots M \}$ based on its weight $W_i(t)$. %(summarizing the queue backlog, direct channel gain and interference channel gains information) and keep sensing the channel until the $m$th %mini-slots.
	
	2.  If contention signal is sensed before the $m$th mini-slot, then the pair $i$ suspends its contention by setting ${\cal I}_i(t) = 0$. %and remains silent in the data %transmission phase of time slot $t$, i.e., 
	
	3. If no contention signal is sensed, the pair $i$ transmits a contention signal in the beginning of the $m$th mini-slot.
	
		- If a collision is detected, the pair $i$ sets ${\cal I}_i(t) = 0$. %i.e., another pair contends for the channel in the $m$th mini-slot, the pair $i$ remains silent during the %time slot $t$ (i.e., ). 
		
		- If no collision is detected, the pair $i$ sets ${\cal I}_i(t) = 1$, and starts its transmission in the beginning of data transmission phase.
	
	4. %If ${\cal I}_i(t) = 1$, the pair $i$ starts transmitting its data at the beginning of data transmission  phase pertaining to time slot $t$. 
	The whole process restarts in the next time slot. 
	\label{alg:CSMA_Based} 
\end{algorithm} 

Based on the contention resolution phase described above, the edge pairs with the smallest backoff time have a chance of earning access rights for the channel. Hence, it is of critical importance to design an efficient association policy mapping small backoff times to the large weights $W_i(t)$ to ensure high utility and to exploit multiuser diversity. Our aim below is to investigate the structure of such efficient policies associating device weights with the mini-slot indices. 

\begin{definition} A mini-slot association policy $\vecbold{\Theta}\paren{\vecbold{w}} = \sqparen{\Theta_1\paren{w_1}, \ldots, \Theta_N\paren{w_N}}$ is a mapping $\vecbold{\Theta}: \Rp^N \mapsto \brparen{1, \ldots, M+1}^N$ such that its $i$th component function $\Theta_i\paren{w_i}$ determines the mini-slot index to which the $i$th pair with weight $w_i$ is assigned.  Further, $\vecbold{\Theta}$ is said to be a threshold policy if all of its component functions can be written as %, where 

\small
\vspace{-0.15in}
\begin{align*}
\Theta_i\paren{w} = \begin{cases} m, & \text{ if } a_m^{(i)} \leq w < a_{m-1}^{(i)}, \ \text{ for } m \in \{1,\ldots,M \} \\
M+1, & \text{ if } w < 0,
\end{cases}
\end{align*}
\normalsize
where $\Theta_i(w) = M+1$ indicates that the edge pair $i$ does not contend for the channel in time slot $t \in \N$. 
%Furthermore, since $W_i(t)$ is %lower bounded by zero and upper bounded by infinity, $a_M^{(i)}(t) = 0$ and $a_0^{(i)}(t) = \infty$ in \eqref{eq:mapping} for all $i$ and %time slot $t$. 
\end{definition}

%$\Theta(W_i(t)) = m_i(t)$. 

We note that the mini-slot index $M+1$ is introduced above for the sake of indicating that transmission from an edge pair is deferred to a next time slot if its weight is negative. In this case, transmissions from such pairs cause more harm to the core AP than its benefit by causing excessive interference. Below, we design a threshold-based mini-slot association policy in which the goal is to operate in close proximity of the optimal point without imposing high complexity as well as providing fairness between edge pairs. In Section \ref{Section: Numerical Results}, we compare the performance of the designed policy with  different threshold-based mini-slot association policies that mainly differ from each other based on how they determine the threshold values.

\vspace{-0.1in} 
\subsection{CADS with Uniform Mapping}

In CADS with uniform mapping, each edge pair is assigned to a mini-slot such that assignment instances are uniformly deployed over all available mini-slots. This is achieved by utilizing the distribution of weights $W_i(t)$ of edge pairs as follows.\footnote{It is assumed that the edge pairs can learn their channel distributions by observing the channel over a period of time \cite{Vaart98}, and the common interference queue backlog, $Z(t)$, is broadcast to the edge devices by the core AP.} %In our model, the fading process is stationary and independent from slot to slot. Hence, the pairs know their instantaneous channel gains, and can learn channel distributions by observing the channel over a period of time \cite{Vaart98}.} 
Let $F_{i,t}(w)$ be the conditional cumulative distribution function (CDF) of $W_i(t)$ at time slot $t \in \N$, defined as  
\begin{align}
F_{i,t}(w) = \Prob{W_i(t) \leq w | W_i(t) \geq 0, Q_i(t), Z(t)}.
\end{align}
%Let $f_{i,t}(w)$ be the corresponding probability density function (PDF). 
Furthermore, let $(F_{i,t})^{-1}(\cdot)$ be the inverse function of $F_{i,t}(\cdot)$. The following lemma indicates how to select the threshold values to achieve uniform distribution over all mini-slot indices.

%Each pair determines the mini-slot in which it will contend by comparing its weight, $W_i(t)$ by the threshold values as illustrated in Fig. \ref{fig:sel_dist}.

\begin{lemma}
	%If $W_i(t) < 0 $, SU $i$ sets ${\cal I}_i(t) = 0$ and do not contend for the channel. 
	%Else it determines the mini-slot as: \\
For time-slot $t \in \N$, consider the mini-slot association policy $\boldsymbol{\Theta}$ defined as 

\vspace{-0.15in}
			\begin{align*}
\Theta_i(w) = \begin{cases} m, & \text{ if } (F_{i,t})^{-1}\left( \frac{M-m}{M}\right) \leq w < (F_{i,t})^{-1}\left( \frac{M-m+1}{M}\right) \\ 
& \hspace{3.2cm} \text{ for } m \in \{1,\ldots,M\} \\
M+1, & \text{ if } w < 0,
\end{cases}
\end{align*}
\label{def:mapping_dist}
for $i \in \brparen{1, \ldots, N}$. Then, for all $t \in \N$, $\boldsymbol{\Theta}$ induces a uniform distribution over mini-slot indices to assign edge pairs.   
\end{lemma}
\begin{IEEEproof}
Directly follows from threshold definitions.
\end{IEEEproof}

The mini-slot association policy above ensures that in the long term, each pair contends in each mini-slot with the same number of times, i.e., on the average with probability of $\frac{1}{M}$ given that its weight is positive. The goal here is to minimize the probability of collision by spreading the contention instances uniformly over all mini-slots. Furthermore, this mini-slot association policy also enforces the scheduling of a good edge device pair with respect to the current channel and queue states. However,  it should be noted that such a uniform mapping policy, although promising, does not necessarily guarantee the scheduling of the best edge pair, i.e., the pair that has the maximum weight, as discussed subsequently.

The first performance loss in the CADS with uniform mapping is due to the contention phase during which an $M\tau$ fraction of whole time slot is used for contention resolution. The second performance loss arises from the possible collisions in the contention phase. Whenever a collision occurs, all edge pairs remain silent during the data transmission phase, and the channel becomes under-utilized. The third  performance loss is the result of imperfect scheduling. The CADS with uniform mapping does not always schedule the edge pair that has the maximum weight. The main underlying reason behind this phenomenon is that each edge device is assigned to a mini-slot uniformly at random with respect to their weights. Although this provides fairness among devices in giving the access rights to the channel (i.e., devices with lower and higher weights are treated equally), it can lead to an assignment of channel access rights to the edge devices with smaller weights.

The next theorem provides an expression for the success probability in the CADS with uniform mapping. 
\begin{theorem} \label{Theorem: CADS Success Probability}
Let $S(t)$ be the event that the contention resolution phase is successful for time slot $t \in \N$ in the CADS with uniform mapping.  Then,
$$ \PRP{S(t)} = \sum_{k=1}^M \frac{N(t)}{M} \paren{\frac{M-k}{M}}^{N(t) - 1},$$
where $N(t)$ is the number of edge pairs with positive weights at time slot $t$. 
\end{theorem}
\begin{IEEEproof}
Please see Appendix B. 
\end{IEEEproof}

We note that the success probability decreases with the number of contending edge users. The worst case scenario is for when $N(t) = N$. Hence, choosing $M$ large with respect to $N$ to guarantee a worst case success probability, we can increase channel utilization. In addition to $\PRP{S(t)}$, another important parameter to assess the efficiency of the CADS with uniform mapping is the weight of the scheduled edge pair for transmission, which is given by $\sum_{i=1}^N{\cal I}_i(t)W_i(t)$. Our simulation results indicate that the CADS with uniform mapping stabilizes the queue sizes around the same stationary points for all users, as expected due to fairness property. In this case, we can obtain a lower bound on the expected scheduled weight with respect to the maximum weight scheduling.        
\begin{theorem}
\label{lemma:bound}
Assume that the edge users observe identically distributed weights over the sample paths generated by the CADS with uniform mapping. For $M \geq N(t)$,   
\begin{eqnarray*}
\E{\sum_{i=1}^N{\cal I}_i(t)W_i(t)}  \geq \alpha(t) \E{W_*(t)}, 
\end{eqnarray*}
where $\alpha(t) = \frac{1-\paren{\frac{N(t)}{M}}^{N(t)}}{\paren{1+\frac{1}{N(t) -1}}^{N(t)-1}}$, and $N(t)$ and $W_*(t)$ are the number of edge pairs with positive weights and the maximum weight at time slot $t \in \N$.  
\end{theorem}
\begin{IEEEproof} Please see Appendix B. 
\end{IEEEproof} 

We note that this is a rather conservative lower bound on $\E{\sum_{i=1}^N{\cal I}_i(t)W_i(t)}$. One reason is that we designed it to be independent of the fading distributions and network states. It can be improved for specific distributions. This bound becomes tighter for $N(t)$ small and $M$ large. Especially, for $M \geq \paren{1+\epsilon} N(t)$, $\epsilon > 0$ and $N(t)$ large, we can write 
$$ \E{\sum_{i=1}^N{\cal I}_i(t)W_i(t)} \geq e^{-1} \E{W_*(t)}.$$
One appealing feature of Theorem \ref{lemma:bound} is that it can help us to relate the scheduled weight in the CADS with uniform mapping to the maximum weight $W^*(t)$ scheduling achieved by the centralized algorithm. In particular, $W_*(t)$ is equal to $\paren{1-M\tau} W^*(t)$ whenever both centralized and distributed algorithms observe the same queue states.\footnote{We assume that $P$ stays the same for the transmission phase so that the total interference energy accumulated at the core AP is scaled accordingly, and the interference due to contention is negligible.} However, the frequency at which they hit the same states is not the same. Hence, the relationship between $\E{W_*(t)}$ and $\E{W^*(t)}$ is more involved. The following theorem provides a lower bound on the performance achieved by the CADS with uniform mapping by considering above observations.     
%Lemma \ref{lemma:bound} indicates that the sum of average weights achieved by the uniform mini-slot association policy is larger than a fraction of the maximum weight. By using Lemma \ref{lemma:bound}, we next characterize the performance of the CADS with uniform mapping.

 \begin{theorem}
\label{thm:optimalcontrol_dist_sel}
Let $\alpha^* = \beta \frac{1-\paren{\frac{N}{M}}^{N}}{\paren{1+\frac{1}{N -1}}^{N-1}}$, $\beta = \frac{\E{W_*(t)}}{\E{W^*(t)}}$ and $M \geq N$.  Suppose $\boldsymbol{x^*} = [x_1^*, \ldots, x_N^*]$ is the average arrival rates produced by the CADS with uniform mapping. Then, for any flow parameter  $V > 0$, the algorithm achieves the following performance bound:
 \begin{align*}
    \sum_{i=1}^N  U_i(x_i^*) &\geq \alpha^* U^* - \frac{B_2}{V} 
		\end{align*}
while bounding the long-term expected queue lengths as:	
		\begin{align*}
    \limsup_{T \rightarrow \infty} \frac{1}{T} \sum_{\tau =
    0}^{T-1}\sum_{i=1}^N \E{Q_i(\tau)} &\leq
    \frac{B_2+V\kappa}{\epsilon_3},
 \end{align*}
 where $B_2,\epsilon_3, \kappa > 0$ are constants and $U^*$ is the optimal aggregate utility of the problem in
 \eqref{eq:opt-objective}-\eqref{eq:const-stability}. %and
 %$\bar{U}$ is the maximum possible aggregate utility.

%This theorem shows that it is possible to get arbitrarily close to
%the optimal utility by choosing $V$ sufficiently large at the
%expense of proportionally increased average queue sizes.
\end{theorem}

\begin{IEEEproof} The proof follows from using the worst case bound in Theorem \ref{lemma:bound} and Corollary 5.2 in \cite{Georgiadis}. \end{IEEEproof} %The proof is given in Appendix \ref{proof:interference_regulated_sel}. \end{IEEEproof} 

%\textit{Remark:} %Assuming that $\tau$ is negligibly small, 

  }
{\allowdisplaybreaks\vspace{-0.1in}
\section{Numerical Results} \label{Section: Numerical Results}

In this section, we present an extensive numerical and simulation study illustrating the analytical results obtained above. 

\subsection{Distributed Schedulers for Performance Comparison}
First, we introduce other CADS algorithms for comparing the performance of our baseline CADS algorithm, which is the CADS with uniform mapping. 

%We consider a two-tier edge network in which edge device pairs are communicating with each other while causing interference to a core AP. Furthermore, 

\subsubsection{CADS with Optimal Weight Mapping}

%In previous subsection, we designed distributed algorithm with uniform mapping, which minimizes collision instances in each mini-slot. %However, in our problem, minimizing collisions does not correspond to optimized performance. This is because, collisions in the first %mini-slots corresponds to larger loss in the performance compared to collisions in later mini-slots, since in the first mini-slot, only pairs %with larger value of $W_i(t)$ contend. Hence, 
This scheduler uses a threshold association policy that maximizes the expected weight of edge device pairs. That is to say, the sequence of threshold values $\brparen{a_m}_{m=1}^M$ is determined as a solution of the following optimization problem

%\small
\vspace{-0.15in}
\begin{align*}
 \max_{\brparen{a_m}_{m=1}^M} \ &\E{{\cal I}_i(t) W_i(t)} = \sum_{m=1}^M \E{{\cal I}_i(t) W_i(t)| \Theta_i(W_i(t)) = m} %\\
%&= \sum_{m=1}^M \E{W_i(t)| \Theta(W_i(t)) = m}\E{{\cal I}_i(t)| \Theta(W_i(t)) = m}  \\
%&  = \int_{a_1}^\infty \left( x f(x) dx \right)\left(1-F(a_1)\right)^{N-1} (1-F(a_1)) \\
%&+\int_{a_2}^{a_1} \left( x f(x) dx \right)\left(1- F(a_2)\right)^{N-1}\left( F(a_1) - F(a_2) \right)\\
%&+\ldots \\
%&+\int_{0}^{a_{M-1}} \left( x f(x) dx \right) \left( F(a_{M-1}) \right)
\end{align*}
for all $i \in \brparen{1, \ldots, N}$. Since the above optimization problem is highly non-linear and dependent on the distributions of fading processes, we are not able to obtain a closed-form solution. Hence, we numerically solve the above problem. Since the CADS with optimal weight mapping maximizes $\E{{\cal I}_i(t) W_i(t)}$ for all $i \in \brparen{1, \ldots, N}$, it results in better performance than that obtained by the uniform mapping, albeit at the expense of increased complexity.

\subsubsection{CADS with Linear Mapping}

%In the previous subsections, we assume that each communication pairs obtains the distribution function and based on this distribution function maps its %weight in slot $t$ into the number of mini-slot in which it contends. This requires the knowledge of distribution of its channel gains, and %calculating mapping function. For some of the nodes with limited power and memory, it may not possible to calculate this mapping function, %e.g., wireless sensors. 

This scheduler uses an easy-to-implement and efficient algorithm for the edge devices having limited power and memory.  Specifically, it utilizes a discrete linear mapping function $\Theta_i$ defined as 
%\begin{definition}
%In the distributed algorithm with linear mapping, the mapping function, $\Theta$, is defined as follows:
			\begin{align*}
\Theta_i(w) = \begin{cases} m, & \text{ if } \frac{(M-m-1)W_{max}}{M}  \leq w < \frac{(M-m)W_{max}}{M}, \\ 
&\ \ \ \ \ \ \ \ \ \ \ \ \ \ \ \ \ \ \ \ \text{ for } m \in \{1,\ldots,M \} \\
M+1, & \text{ if } w < 0,
\end{cases}
%\label{eq:mapping}
\end{align*}
%	\label{def:mapping_dist}
for all $i \in \brparen{1, \ldots, N}$, where $W_{max}$ is a large constant representing an upper bound on the weight realizations. All edge devices agree on the value of $W_{max}$ and then use the above mapping to determine mini-slot indices.
%\end{definition}
 
%We should note that the performance of the algorithm with linear mapping mainly depends on two factors: (1) The selection of $W_{max}$. That %is to say, if $W_{max}$ is too small, then the contentions performed by communication pairs, get clustered around the first mini-slots. This will %result high volume of collisions which decreases the performance of the algorithm. On the contrary, if it is too large, then users tend to %select later mini-slots to contend, and this again causes high volume of collisions and degrades the performance of the algorithm. Thus the %selection of $W_{max}$ plays the crucial role for the performance of the algorithm. (2) The shape of actual distribution function of the %weights. Since we assume linear mapping, if the shape of the distribution is closer to a linear function (or the error between actual %function and linear approximation of the function is small), then we can say that the algorithm performs well. Note that, any other mapping %function can be used instead of linear mapping based on the shape of the distribution function. However, linear function is simple %approximation and does not require high power or memory for calculation. Furthermore, it works well for a wide range of distribution %function, e.g., Rayleigh or Nakagami fading.  
\subsubsection{Interference Regulated Distributed Scheduler}
We modify the baseline algorithm proposed in \cite{Xue13} to obtain an interference regulated distributed scheduler (IRDS). 

%regulated queue model to consider the interference constraint, and it is called Interference Regulated Distributed Scheduler (IRDS), which is based on the baseline algorithm in . 

The operation of IRDS is divided into two phases: (i) contention phase and (ii) data transmission phase, similar to the CADS algorithms above. To facilitate the discussion, we introduce two new random variables related to contention and scheduling phases. The first one is the contention variable, $a_i(t)$, that is 1 with probability $\frac{1}{N}$, and 0 with probability $\frac{N-1}{N}$. 
%has a probability distribution given by
%\begin{align}
%a_i(t)=\begin{cases} 1, & \text{ w.p.  } \frac{1}{N} \\
% 0, & \text{ w.p.  } \frac{N-1}{N}.
%\end{cases}
%\end{align}
The second one is the transmission variable, $p_i(t)$, that is 1 with probability  $\frac{e^{W_i(t)}}{e^{W_i(t)}+1}$, and 0 with probability $\frac{1}{e^{W_i(t)}+1}$, %independent over pairs, and has distribution given by
%\begin{align}
%p_i(t)=\begin{cases} 1, & \text{ w.p.  } \frac{e^{W_i(t)}}{e^{W_i(t)}+1} \\
% 0, & \text{ w.p.  }  \frac{1}{e^{W_i(t)}+1},
%\end{cases}
%\end{align}
where $W_i(t)$ is the weight of edge pair $i$ defined in \eqref{eq:weight} for all $i \in \brparen{1, \ldots, N}$. %The dynamics of
%real queue, $Q_i(t)$, and virtual queue, $Z(t)$ are as in
%\eqref{eq:real_queue} and \eqref{eq:interference_queue}. 
We note that the transmission variable takes also into account interference level that is caused to the core AP.  This is the reason why the algorithm is regulated with respect to the interference level.

The scheduling decision of edge pair $i \in \brparen{1, \ldots, N}$ depends on the following three conditions: 

Condition (1): The contention of pair $i$  is successful, i.e.,
$a_i(t)\prod_{j \neq i}(1-a_j(t))= 1$.

Condition (2): None of the neighboring pairs were scheduled in the
previous time slot, i.e., $\sum_{j \neq i} {\cal I}_j(t-1) = 0$

Condition (3): The transmission variable $p_i(t) = 1$. 
 
Based on these three conditions, the scheduling phase consists of three different cases, as given by Algorithm \ref{alg:int_reg}. 
\begin{algorithm}
\caption{Scheduling Phase of IRDS}   
 Each edge pair determines ${\cal I}_i(t)$ according to 

Case 1: ${\cal I}_i(t) = 1$ if the conditions (1), (2) and (3) hold.

Case 2: If the condition (1) does not hold and the condition (3) holds, then
${\cal I}_i(t) = {\cal I}_i(t-1)$.

Case 3: Otherwise, ${\cal I}_i(t) = 0$.

\label{alg:int_reg}
\end{algorithm}

Notice that the IRDS only considers a single round contention unlike the proposed CADS algorithms, which limits the edge network performance as shown in simulation results.

 \begin{figure} 
\centerline{{\includegraphics[width=3.0in]{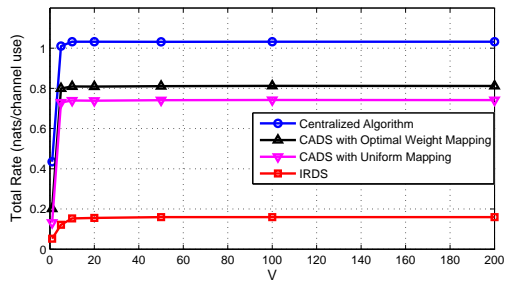}}}
\caption{Performance of the edge network as a function of $V$. Both centralized and distributed dynamic control mechanisms are considered.}
\label{fig:D2D_V}
\end{figure}

\subsection{Simulation Results}

In the simulations below, we consider iid Rayleigh fading channels, with direct and interference power gains given by exponential distributions having
means $2$ and $1$, respectively. The service rate functions $R_i(t)$ are chosen to be $R_i(t) = \log\left(1+\frac{Ph_i(t)}{I_i(t)+1}\right)$ for all $i \in \brparen{1, \ldots, N}$. Further, we use logarithmic utility functions to measure edge device satisfaction for an achieved rate value $x_i$ for $i=1, \ldots, N$. Specifically, the edge pair $i$ obtains a proportionally fair utility of $\log(1+x_i)$ at rate $x_i$. Since the sum of utility functions are taken as the objective function to be maximized, we obtain the utility maximizing arrival rates of edge devices, and the rates depicted in the graphs below are the sum of these optimum arrival rates.  To simulate $I_i(t)$, we generate $20$ exponentially distributed interfering links to the edge network devices, and the mean channel gains of these links are randomly chosen  between $0.1$ and $0.3$. The total interference at the $i$th edge device pair is the weighted sum of the gains of these interfering links, with weights being the transmission powers that are set to unity. Unless otherwise stated, the number of edge device links and mini-slots are set to $100$ and $200$, respectively.  Furthermore, we take $\gamma=0.1$ and $\tau = 10^{-4}$ except simulations conducted with respect to these parameters. We will only consider average interference constraints in the simulations below for the sake of having a clean exposition. 

%We consider logarithmic utility functions for achieving proportional fairness in our simulations below. Specifically, the edge pair $i$ obtains a utility of $\log(1+x_i)$ at rate $x_i$. %\footnote{We utilize logarithmic utility function to provide proportional fairness.}.  
%The rates depicted in the graphs are the sum of arrival rates of all devices and the unit of the plotted rates is natts/channel use. First, the main channel between communication pairs and interference channel between transmitters of pairs are modeled as i.i.d. Rayleigh fading Gaussian channels, and $R_i(t) = \log\left(1+\frac{Ph_i(t)}{I_i(t)+1}\right)$. Thus, the main and interference power gains are exponentially distributed with means 2 and 1, respectively. The noise normalized power is $P=1$. { \color{red}In addition, we select 20 interfering links to the edge network devices, and the mean channel gains of these links are randomly chosen  between 0.1 and 0.3 from exponential distribution. The total interference to the edge pair $i$ in time slot $t$, $I_i(t)$, is the sum of all interfering links' gains.} Furthermore, in experiments, we only consider average interference constraint. We compare the performance of our algorithm with different mini-slot association policies and that of widely used regulated queue approach.

We compare the performance curves for CADS with uniform mapping, CADS with optimal weight mapping, CADS with linear mapping and IRDS with the performance obtained through the centralized algorithm. To start with, we investigate the effect of the system parameter $V>0$ on our dynamic control algorithms in Fig. 5.  As expected, the total rate of all algorithms increases with increasing $V$ values, and Fig. \ref{fig:D2D_V} shows that the rate achieved by the centralized algorithm for $V \geq 20$ converges to its optimal value fairly closely verifying the results in Theorem \ref{thm:optimalcontrol}. 

Furthermore, the distributed algorithm achieving the best performance is the CADS with optimal weight mapping. It attains an average rate over $80 \%$ of the total rate of the centralized algorithm. The CADS with uniform mapping exhibits a performance curve fairly close to the one with optimal weight mapping, achieving around  $70 \%$ of the total rate of the centralized algorithm. 
%The reason for such slightly worse performance is that even though the CADS with uniform mapping minimizes collisions in each mini-slot, minimizing collisions does not correspond to an optimized performance. This is because collisions in the earlier mini-slots correspond to a larger loss in the performance compared to those in later mini-slots.  
Based on the derived success probability distribution in Theorem \ref{Theorem: CADS Success Probability}, we observe that the success in the earlier mini-slots become more likely than those in the later mini-slots, which is eventually more beneficial due to higher weight scheduling.  However, such collision minimization does not correspond to an optimized performance with max-weight scheduling since it is still possible that the successfully scheduled edge pair does not have the maximum weight. The first factor puts an upward pressure to increase the performance of the CADS with uniform mapping, whereas the second one puts a downward pressure to decrease the performance of the CADS with uniform mapping. At the end, they balance each other, leading to an observed slightly worse performance of the CADS with uniform mapping.   

Among four distributed schedulers, the IRDS has the worst performance achieving only approximately $30 \%$ of that of the centralized algorithm. There are mainly two reasons about such a poor performance for the IRDS. Firstly, the IRDS cannot fully take advantage of channel diversity due to using single contention period. It schedules an edge device randomly, and then decides whether the scheduled edge device should transmit or not after the scheduling decision. Secondly, to schedule a new edge device, the algorithm always requires an idle time-slot, and this increases the number of idle time-slots leading to under-utilization of the channel.

\begin{figure*}
\centerline{ \subfloat[Performance with respect to $\gamma$]{\includegraphics[width=3.0in]{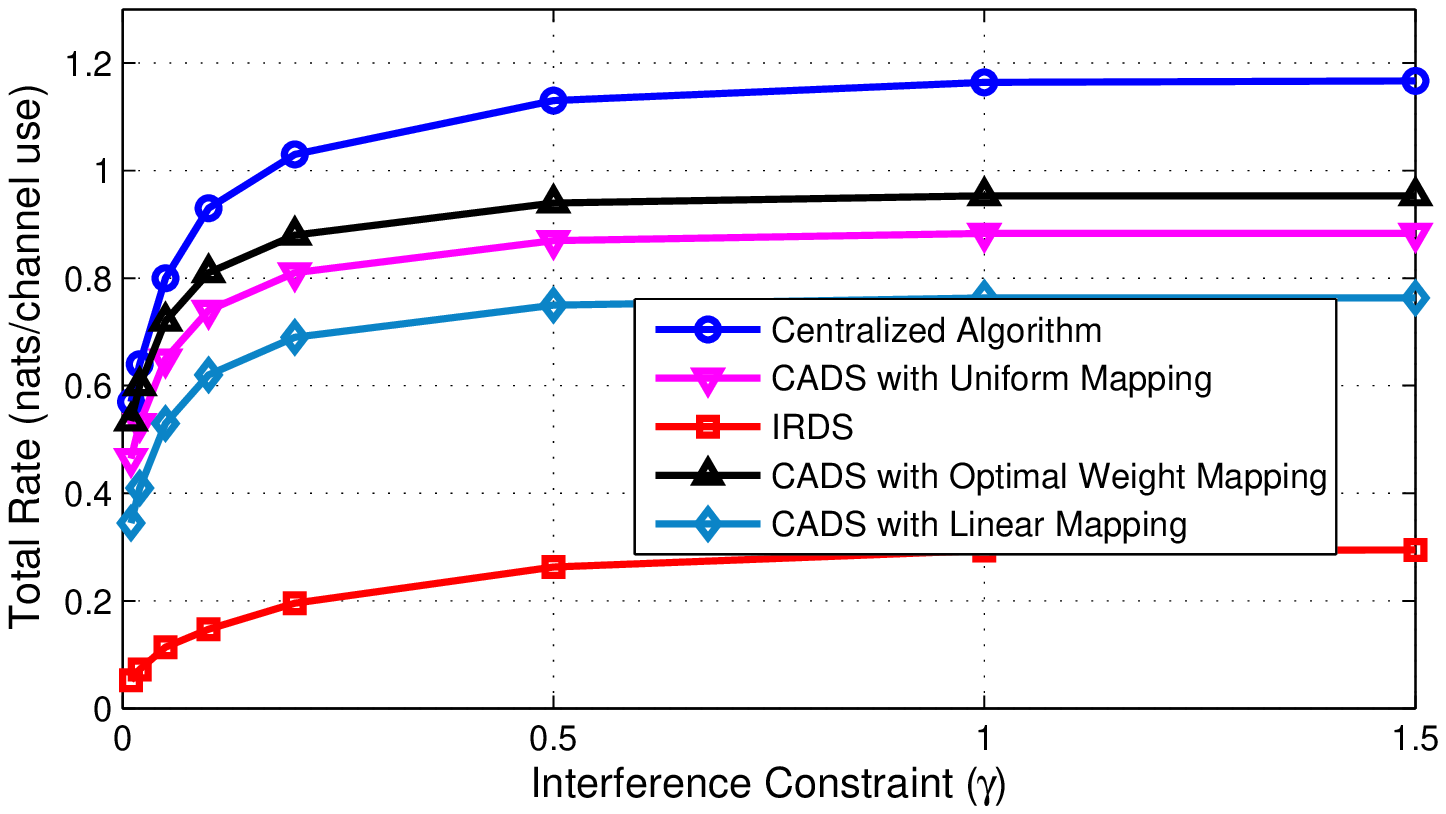}
\label{fig:D2D_gamma}} \subfloat[Performance with respect to $N$]{\includegraphics[width=3.0in]{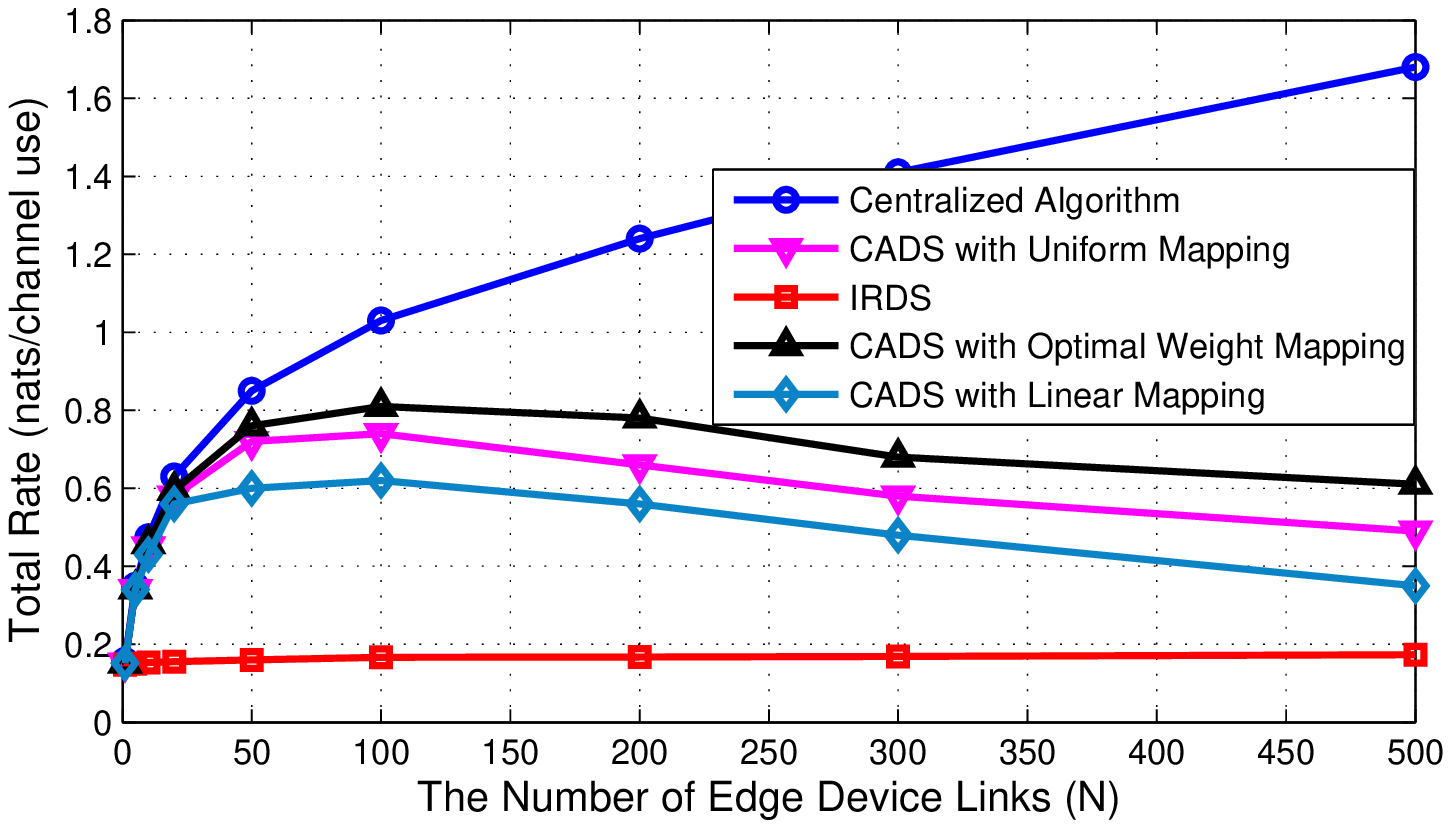}%
\label{fig:numnode}  }} \caption{Performance of the edge network as a function of $\gamma$ and $N$. Both centralized and distributed dynamic control mechanisms are considered.}
\end{figure*}

For the rest of the experiments, we take $V= 100$. In Figs. \ref{fig:D2D_gamma} and \ref{fig:numnode}, we analyze the effect $\gamma$ and $N$ on the system performance, respectively. As illustrated in Fig. \ref{fig:D2D_gamma}, the total rate of all algorithms increases with increasing $\gamma$. This is because for low $\gamma$ values, in order to satisfy a tight interference constraint, a larger fraction of time-slots remains idle, i.e., smaller number of transmission opportunities are given to edge device pairs.  Starting around $\gamma$ = 0.5, the interference constraint becomes inactive. In Fig. \ref{fig:numnode}, we first notice that the performance of the IRDS does not change with increasing number of pairs and only achieves $90\%$ of the rate achieved by the centralized algorithm when the number of device pairs is equal to one. This result arises from the fact that the IRDS cannot take the advantage of diversity gains available in fading channels. On the other hand, the CADS algorithms achieve performances that are closer to the centralized algorithm due to taking advantage of link diversity. As expected, the CADS with optimal weight mapping performs the best, whereas the one with linear mapping has the worst performance. In the case of linear mapping, we sacrifice some performance in favor of reducing mapping complexity. However, as $N$ increases, the collisions in contention phase become more dominant compared to diversity gain, and as $N>500$, the total rate of CADS algorithms decreases. 

We should note that the performance of the CADS with linear mapping mainly depends on two factors. The first one is the selection of $W_{max}$.  If $W_{max}$ is too small, then the channel access attempts get clustered around the earlier mini-slots. This results in high volume of collisions and decreases the performance of the algorithm. On the contrary, if $W_{max}$ is too large, then the edge devices tend to select later mini-slots to contend. This again causes high volume of collisions. In the light of this discussion, we select $W_{max}$ such that the worst case probability of weights being larger than $W_{max}$ is $\frac{1}{M}$. The second factor that affects the performance of the CADS with linear mapping is the shape of the CDFs of the weights. If the CDFs are closer to linear functions, then we can say that the algorithm can perform well. 

%Furthermore, as the number of edge pairs increases, the collisions during the contention phase increases. This results in an increase in the difference between the performance of centralized scheduler and CADS algorithms.

\begin{figure*}
\centerline{ \subfloat[$\tau = 10^{-4}$]{\includegraphics[width=3.0in]{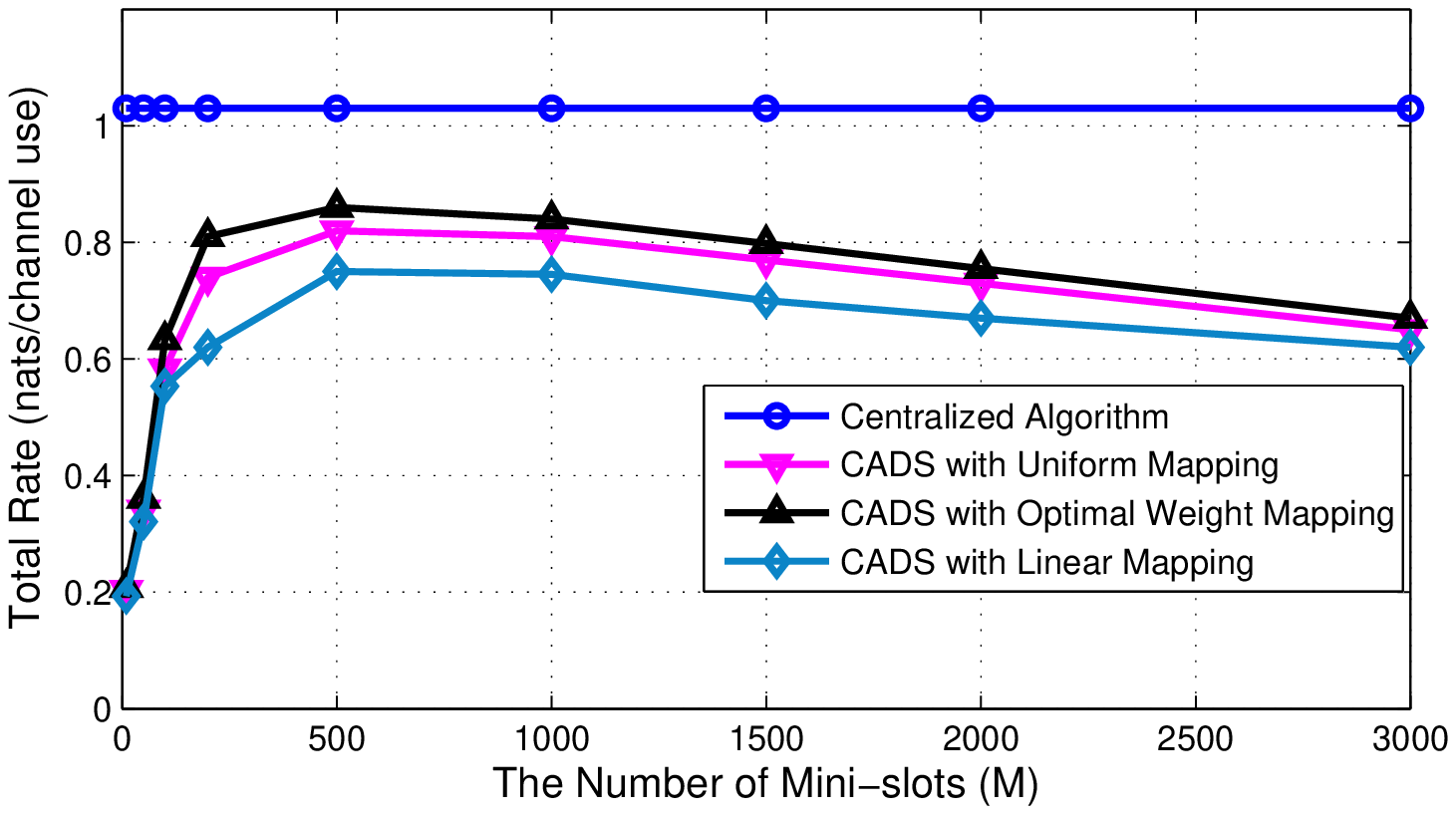}
\label{fig:M1}} \subfloat[$\tau = 2\cdot10^{-4}$]{\includegraphics[width=3.0in]{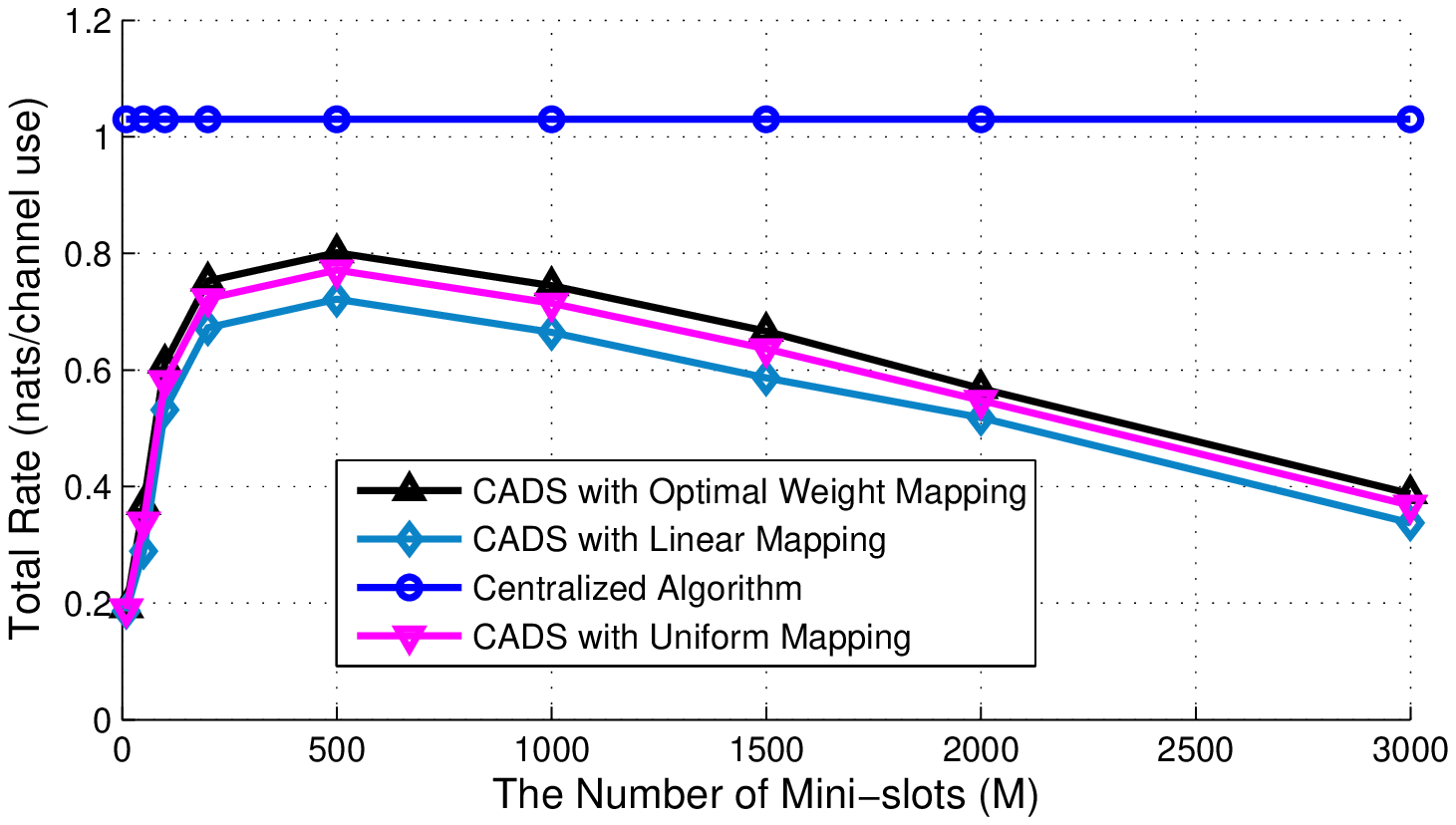}%
\label{fig:M2}  }} \caption{Performance of the edge network as a function of $M$. Both centralized and distributed dynamic control mechanisms are considered.}
\end{figure*}

Next, we analyze the effect of the number of mini-slots on the edge network performance. In Fig. \ref{fig:M1}, we take $\tau = 10^{-4}$ and in Fig. \ref{fig:M2}, we take $\tau = 2\cdot 10^{-4}$. As illustrated in these figures, the total data rate increases initially with increasing values of $M$. This is due to the fact the edge network experiences less collisions with increasing $M$. However,
when $M$ is excessively high, the emphasis on reducing collisions becomes less significant, but the loss due to the contention window size is more prominent. As a result, the performance of CADS algorithms gets worse. %The optimal $M$ is achieved when $M$ is around $600$ in Fig. \ref{fig:M1}. In Fig. \ref{fig:M2}, we notice that the optimal value of $M$ shifts to the left, which is now around 400.
Furthermore, the decrease of the total rate due to having long contention phase is sharper in Fig. \ref{fig:M2}. This is because higher $\tau$ corresponds to higher cost of implementing mini-slots, so the optimal $M$ decreases.

\begin{figure*}
\centerline{ \subfloat[Performance with respect to $\gamma$]{\includegraphics[width=3.0in]{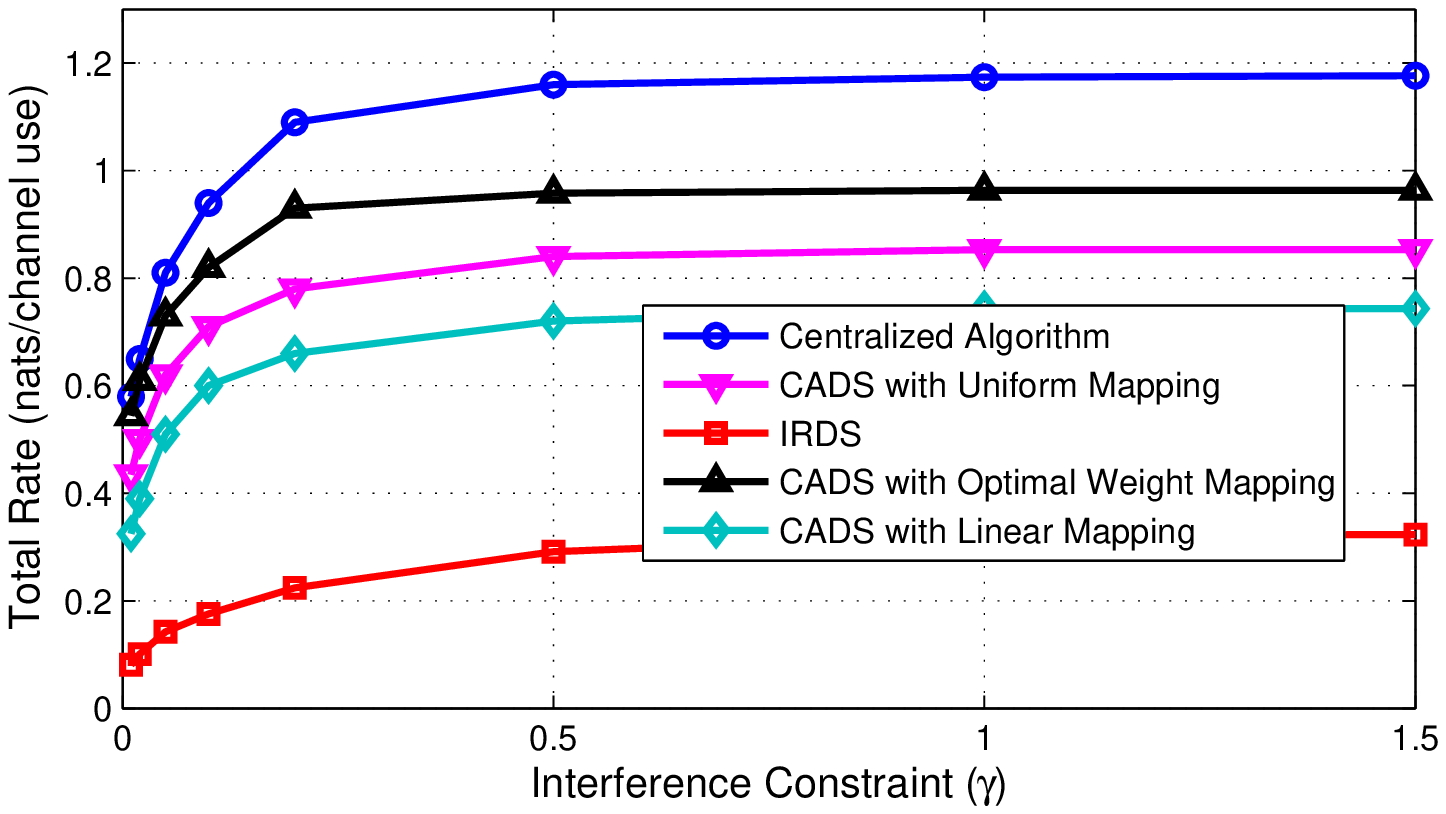}
\label{fig:gamma_NU}} \subfloat[Performance with respect to N]{\includegraphics[width=3.0in]{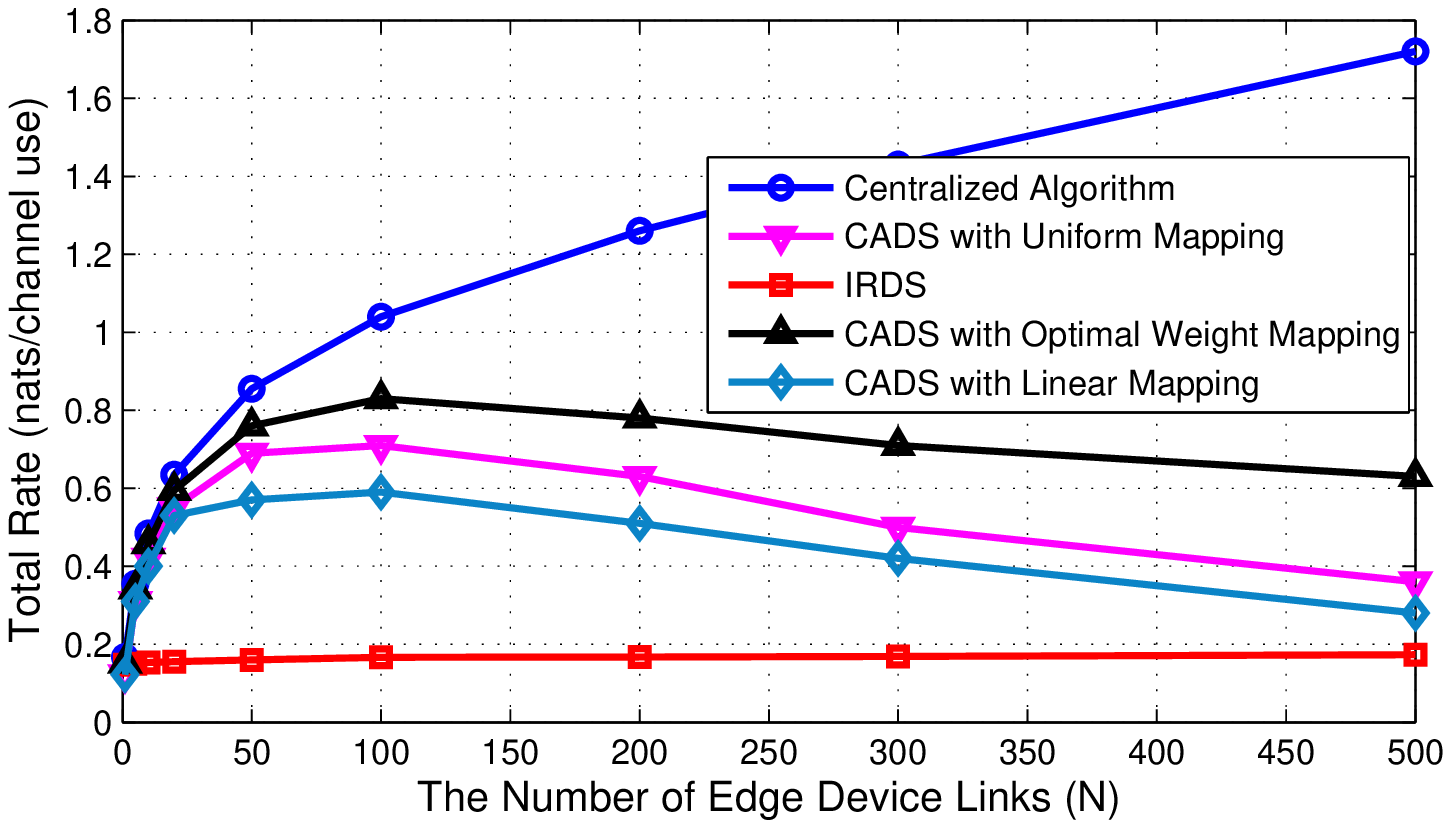}%
\label{fig:numnode_NU}  }} \caption{Performance of the edge network as a function of $\gamma$ and $N$ with non-iid channels. Both centralized and distributed dynamic control mechanisms are considered.}
\end{figure*}

Lastly, we investigate the edge network performance with respect to $\gamma$ and $N$ when the channels are not iid. Precisely, the direct and interference channel gains of edge pairs are chosen uniformly at random from the intervals $\sqparen{1.2, 2.8}$ and $\sqparen{0.2,1.8}$, respectively. %In addition, we take $10$ runs and the rates depicted in Fig. \ref{fig:gamma_NU} and \ref{fig:numnode_NU} are the average of these 10 runs. From Fig. \ref{fig:gamma_NU} and \ref{fig:numnode_NU}, 
The same observations above continue to hold. Differently, we notice that the CADS with uniform mapping perform slightly worse than the case when the channel gains are iid. %, e.g., %the algorithm with optimal weight mapping now achieves approximately $85 \%$ of centralized algorithm whereas it achieves $90 \%$ of %centralized algorithm when the channel gains are iid. 
The reason is that when the channel gains are non-iid, the difference between the weights and the queue sizes of edge devices becomes larger. Then, the CADS algorithms cannot schedule the edge pair with the highest weight more frequently due to having different mapping intervals at each pair. }
{\allowdisplaybreaks\section{Conclusion}

In this paper, we considered the problem of cross-layer design for underlay edge networks. We first derived the stability region for the edge network subject to interference constraints at the core AP. We then designed a cross-layer flow control and scheduling algorithm that maximizes the total collective utility of the edge network within the stability region. This is a centralized algorithm that requires knowledge of edge device queue sizes for scheduling. We also studied distributed implementation issues by relaxing  the design of the centralized algorithm. The loss from distributed operation was characterized. Finally, an extensive simulation and numerical study was performed. The derived analytical results as well as the performance gap between the centralized and distributed schedulers were illustrated as a function of various edge network parameters. As a future work, we plan to investigate the performance of interference-aware mobile edge network in a multi-user scheduling setting where a number of edge devices can transmit in the same time-slot.

%resource allocation in wireless two-tier network where edge network device pairs have information to be shared.  However, the interference to the core AP should be kept arbitrarily low caused by the transmission between edge device pairs. First, we studied the stability region of such interference-aware network, and compare the region to the case when there is no interference constraint. Then, we described a cross-layer dynamic algorithm, and we proved that our algorithm achieves utility arbitrarily close to achievable optimal utility. 
}
%\tiny
\bibliographystyle{IEEEtran}
\bibliography{d2dcom}

\appendices

{\allowdisplaybreaks
\vspace{-0.0in}
\section{Proof of Theorem 1}
\label{proof:optimalscheduling}

Assume that the primal optimization problem in \eqref{eq:obj_linear}-\eqref{eq:sch_const} is feasible with an interior point. Let $\Omega \subseteq \Psi$ contain all scheduling policies satisfying $\sum_{i=1}^N  {\cal I}_i\paren{\vecbold{h}, \vecbold{g}, \vecbold{I}} \leq 1$ and ${\cal I}_i\paren{\vecbold{h}, \vecbold{g}, \vecbold{I}} = 0$ if $P g_i > \nu$ for all fading and interference states. Then, the primal problem can be recast as 
\begin{eqnarray}
\begin{array}{ll}
\mbox{maximize}_{\vecbold{\cal I} \in \Omega} & \E{{\cal I}_i (\boldsymbol{h},\boldsymbol{g},\boldsymbol{I}) R_i} \\
\mbox{subject to} & \E{{\cal
I}_j(\boldsymbol{h},\boldsymbol{g},\boldsymbol{I})R_j} \geq \alpha_j, \forall j \neq i \\
& \E{\sum_{j=1}^N
P{\cal I}_j(\boldsymbol{h},\boldsymbol{g},\boldsymbol{I}) g_j} \leq \gamma
\end{array}. \label{Eqn: Primal Optimum}
\end{eqnarray} 

Consider the  Lagrangian for the above problem defined as
\begin{eqnarray*}
L\paren{\vecbold{\cal I}, \vecbold{\lambda}, \mu} = \E{{\cal I}_i (\boldsymbol{h},\boldsymbol{g},\boldsymbol{I}) R_i} + \sum_{j \neq i} \lambda_j\paren{\E{{\cal
I}_j(\boldsymbol{h},\boldsymbol{g},\boldsymbol{I})R_j} - \alpha_j} \\ \hspace{0cm} + \mu \paren{\gamma - \sum_{j=1}^N \E{P{\cal I}_j(\boldsymbol{h},\boldsymbol{g},\boldsymbol{I}) g_j}},
\end{eqnarray*}
where $\mu \geq 0$ and $\vecbold{\lambda} \in \Rp^{N-1}$ are associated Lagrange multipliers \cite{Luenberger}.   
We observe that the set $\Omega$ is convex in the sense that if $\vecbold{\cal I}_1 \in \Omega$ and $\vecbold{\cal I}_2 \in \Omega$, then $a \vecbold{\cal I}_1 + \paren{1-a}\vecbold{\cal I}_2$ also lies in $\Omega$. Hence, there exists $\mu^*$ and $\vecbold{\lambda}^*$ such that the optimal value of the following convex problem 
\begin{eqnarray}
\max_{\vecbold{\cal I} \in \Omega}  L\paren{\vecbold{\cal I}, \vecbold{\lambda}^*, \mu^*} \label{Eqn: Dual Optimum}
\end{eqnarray}
coincides with the optimal value of the primal problem \cite{Luenberger}. This problem is easy to solve since the constraints defining $\Omega$ are given for each state. Therefore, \eqref{Eqn: Dual Optimum} can be solved for each fading and interference state, which leads to 
\begin{eqnarray}
\setlength{\nulldelimiterspace}{0pt}
{\cal I}^*_j\paren{\vecbold{h}, \vecbold{g}, \vecbold{I}} = \left\{\begin{IEEEeqnarraybox}[\relax][c]{l's}
1, & if $j \in \argmax_{k \in C} W_k$\\
0, & otherwise
\end{IEEEeqnarraybox}\right. \label{Eqn: Centralized Scheduler2}
\end{eqnarray}
for all $j \in \brparen{1, \ldots, N}$, where $C = \brparen{j: P g_j \leq \nu \mbox{ and } W_j \geq 0}$ and $W_j$ is given by $W_j = \lambda_j^* R_j - \mu^* P g_j$ for $j \neq i$ and $W_i = R_i - \mu^* P g_i$.  It also holds that any solution for \eqref{Eqn: Primal Optimum} is also a solution for \eqref{Eqn: Dual Optimum}. This concludes the proof since the above solution given in \eqref{Eqn: Centralized Scheduler2} is unique almost surely.  

}
 {\allowdisplaybreaks\vspace{-0.0in}
\section{Performance of the CADS with Uniform Mapping}
\label{proof:bound_weight}

\subsection{Proof of Theorem \ref{Theorem: CADS Success Probability}}

Assume there are $N(t)$ edge pairs with positive weights at time slot $t \in \N$. Without loss of generality, label them as $W_i$ for $i=1, \ldots, N(t)$. Let $m_i = \Theta\paren{W_i}$ and $m^* = \min_{1 \leq i \leq N(t)} m_i$.   Observe that  $m^*$ is the minimum of $N(t)$ uniformly and independently distributed random variables over $\brparen{1, \ldots, M}$. Then,
\begin{eqnarray*}
\PRP{S(t) \mbox{ and } m^*=m} = \frac{N(t)}{M}\paren{\frac{M-m}{M}}^{N(t)-1}
\end{eqnarray*}
since at least one edge pair must be assigned to the mini-slot $m$ and the rest must be assigned to those with higher indices for both $S(t)$ and $m^*=m$ to hold correct. Summing over $m$, we obtain $\PRP{S(t)}$ as stated in the theorem.   
\subsection{Proof of Theorem \ref{lemma:bound}}

%In this Appendix, our aim is to obtain a bound on the average weight achieved by CADS with uniform mapping with respect to the weight achieved by the centralized algorithm given in Section \ref{control}, i.e., the max weigh. Also, we define the maximum weight, 
%Let $W^*(t) = \max_{1\leq i \leq N} W_i(t)$ and 

%Below, we provide the worst case performance analysis. Let $W_i(t)$ be the weight of the edge user $i$ at time slot $t$. Consider the event that all $W_i(t)$'s are larger than zero, which means all edge users contend for channel access. All the probability calculations below will be with respect to the conditional probability measure on this event although we do not show it explicitly. 

Let $\mathcal{F}_t$ be the $\sigma$-algebra generated by the states $\brparen{Q_i(t)}_{i=1}^N$ and $Z(t)$. Using the same notation above, let $W_*(t) = \max_{1\leq i \leq N(t)} W_i$. Observe that we always have $\sum_{i=1}^N {\cal I}_i(t)W_i(t) = W_*(t)$ on the event $S$. Then,  
\begin{eqnarray*}
\E{\sum_{i=1}^N {\cal I}_i(t)W_i(t) \Big \vert \mathcal{F}_t} &=& \E{\sum_{i=1}^N {\cal I}_i(t)W_i(t); S(t) \Big \vert \mathcal{F}_t} \\
&=& \E{W_*(t); S(t) \big \vert \mathcal{F}_t}, %\\
%&=& \E{W^*(t) \big\vert E} \PRP{E} \\
%&\geq&\E{W^*(t)} \PRP{E},
\end{eqnarray*}
where $\E{\sum_{i=1}^N {\cal I}_i(t)W_i(t); S(t) \big \vert \mathcal{F}_t}$ indicates the expectation of the random variable $\sum_{i=1}^N {\cal I}_i(t)W_i(t)$ on the event $S(t)$. The next lemma provides an expression for $\E{W_*(t); S(t) \big \vert \mathcal{F}_t}$.  
\begin{lemma} \label{Lemma: Max Weight Lemma 1}
\begin{eqnarray*}
\E{W_*(t); S(t) \big \vert \mathcal{F}_t} \hspace{8cm}\\
 = N(t) \sum_{m=1}^M \paren{\frac{M-m}{M}}^{N(t)-1} \E{W; a_m \leq W < a_{m-1} \big \vert  \mathcal{F}_t}, \hspace{1.7cm}
\end{eqnarray*}
$W$ is a generic random variable having the same conditional distributions with $W_i$'s and $a_m$'s are the associated threshold values in the CADS with uniform mapping.  
\end{lemma}
\begin{IEEEproof}
Let $E_i$ be the event that $m_i = m$ and $m_j > m$ for all $j \neq i$. Then, we can write 
\begin{eqnarray*}
\E{W_*(t); S(t) \big \vert \mathcal{F}_t} &=& \sum_{m=1}^M \E{W_*(t); S(t), m^* = m \big \vert \mathcal{F}_t} \\
&=& \sum_{m=1}^M \sum_{i=1}^{N(t)} \E{W_*(t); E_i \big \vert \mathcal{F}_t} \\
&=& \sum_{m=1}^M \sum_{i=1}^{N(t)} \E{W_*(t) \big \vert E_i, \mathcal{F}_t} \PRP{ E_i \big \vert \mathcal{F}_t } \\
&=& \sum_{m=1}^M \sum_{i=1}^{N(t)} \E{W_i \big \vert E_i, \mathcal{F}_t} \PRP{E_i \big \vert \mathcal{F}_t}.
\end{eqnarray*}
Since the weights are assumed to be conditionally iid, we have  $\E{W_i \big \vert E_i, \mathcal{F}_t} = M \E{W; a_m \leq W < a_{m-1} \big \vert \mathcal{F}_t}$. Further, $\PRP{E_i \big \vert \mathcal{F}_t} = \frac{1}{M} \paren{\frac{M-m}{M}}^{N(t)-1}$.  Combining these results, we complete the proof of Lemma \ref{Lemma: Max Weight Lemma 1}. 
\end{IEEEproof}

Now, we will bound $\E{W_*(t) \big \vert \mathcal{F}_t}$. First observe that 
\begin{eqnarray*}
\E{W_*(t) \big \vert \mathcal{F}_t} &=& \sum_{m=1}^M \int_{a_m}^{a_{m-1}}N(t) f_W(w)\paren{F_W(w)}^{N(t)-1} dw \\
%&\leq&  N(t) \sum_{m=1}^M \paren{\frac{M-m+1}{M}}^{N(t)-1} \int_{a_m}^{a_{m-1}} f_W(w) dw \\
&\leq& N(t) \sum_{m=1}^M \paren{\frac{M-m+1}{M}}^{N(t)-1} \\ 
 && \hspace{1.8cm} \times \E{W; a_m \leq W < a_{m-1}\big \vert \mathcal{F}_t}, 
\end{eqnarray*}
where $f(w)$ and $F(w)$ are the conditional PDF and CDF for $W$, respectively. Dividing above sum into two parts, we obtain
\begin{eqnarray*}
\E{W_*(t) \big \vert \mathcal{F}_t} \paren{1-N(t) \sum_{m=1}^{N(t)-1} \paren{\frac{m}{M}}^{N(t)-1}} \hspace{8cm} \\
\leq N(t) \sum_{m=1}^{M-N(t)+1} \paren{\frac{M-m+1}{M}}^{N(t)-1} \E{W; a_m \leq W < a_{m-1}\big \vert \mathcal{F}_t}.    \hspace{4.8cm}
\end{eqnarray*}

Now, we compare the terms $\paren{\frac{M-m+1}{M}}^{N(t)-1}$ and $\paren{\frac{M-m}{M}}^{N(t)-1}$. The ratio of these two terms is bounded above by $\paren{\frac{M-m+1}{M-m}}^{N(t) - 1} \leq \paren{1+\frac{1}{M-m}}^{N(t)-1} \leq \paren{1+\frac{1}{N(t) -1}}^{N(t)-1}$. Using Lemma \ref{Lemma: Max Weight Lemma 1}, this shows that 
\begin{eqnarray*}
\E{W_*(t) \big \vert \mathcal{F}_t} \paren{1-N(t) \sum_{m=1}^{N(t)-1} \paren{\frac{m}{M}}^{N(t)-1}} \hspace{8cm} \\
\leq \paren{1+\frac{1}{N(t) -1}}^{N(t)-1} \E{W_*(t); S(t) \big \vert \mathcal{F}_t}.  \hspace{5.5cm}
\end{eqnarray*}
The above sum can be bounded above by $\frac{1}{N(t)} \paren{\frac{N(t)}{M}}^{N(t)}$. Hence, the following bound holds
\begin{eqnarray*}
\E{W_*(t); S(t) \big \vert \mathcal{F}_t} \geq \frac{1-\paren{\frac{N(t)}{M}}^{N(t)}}{\paren{1+\frac{1}{N(t) -1}}^{N(t)-1}} \E{W_*(t) \big \vert \mathcal{F}_t}. 
\end{eqnarray*}
Averaging over $\mathcal{F}_t$, we finally have 
\begin{eqnarray*}
\E{W_*(t); S(t)}  \geq \frac{1-\paren{\frac{N(t)}{M}}^{N(t)}}{\paren{1+\frac{1}{N(t) -1}}^{N(t)-1}} \E{W_*(t)} ,
\end{eqnarray*}
which completes the proof.

\normalsize 
%This concludes our proof.

%Note that, in \eqref{eq:loss_imperfect}, we take into account the loss due to imperfect scheduling and mini-slot implementation. By inserting the inequality in \eqref{eq:ineq_suc_trans} into \eqref{eq:max_weight}, we obtain

%\small
%\begin{align*}
%\E{{\cal I}_i(t)W_i(t)} &\geq (1-M\tau)(1-\beta) \sum_{m=1}^M\E{W^*(t)\big\vert {\cal I}_i(t) = 1 ,\Theta(W^*(t)) = m} \\
%&\ \ \ \ \ \ \ \ \Prob{\Theta(W^*(t)) = m} \frac{N}{M}\left(1-\frac{1}{M}\right)^{N-1} \\
%&=(1-M\tau)(1-\beta)\E{W^*(t)}\frac{N}{M}\left(1-\frac{1}{M}\right)^{N-1},
%\end{align*}
%where \\
%\footnotesize
%$\E{W^*(t)} =	\sum_{m=1}^M\E{W^*(t)\big\vert {\cal I}_i(t) = 1 ,\Theta(W^*(t)) = m}\Prob{\Theta(W^*(t)) = m}	$. \normalsize 

%This concludes our proof.

%%\input{appendix_dist_sel}
}

\end{document}